\documentclass[%
reprint,
amsmath,amssymb,
aps,
pra,
]{revtex4-2}
\usepackage{graphicx}
\usepackage{dcolumn}
\usepackage{bm}

\usepackage{xcolor} 
\usepackage{ulem} 

\begin{document}
	\preprint{APS/123-QED}

\title{Excess noise and photo-induced effects in highly reflective crystalline mirror coatings}

\author{Jialiang Yu}
\email[e-mail: ]{jialiang.yu@ptb.de}
\author{Sebastian H\"afner}
\author{Thomas Legero}
\author{Sofia Herbers}
\author{Daniele Nicolodi}
\author{Chun Yu Ma}
\author{Fritz Riehle}
\author{Uwe Sterr}
\email[e-mail: ]{uwe.sterr@ptb.de}
\affiliation{Physikalisch-Technische Bundesanstalt, Bundesallee 100, 38116 Braunschweig, Germany}
\author{Dhruv Kedar}
\author{John M. Robinson}
\author{Eric Oelker}
\altaffiliation[current address: ]{University of Glasgow, Institute for Gravitational Research, School of Physics and Astronomy, Glasgow G12 8QQ, United Kingdom}
\author{Jun Ye}
\email[e-mail: ]{Ye@jila.colorado.edu}
\affiliation{JILA, National Institute of Standards and Technology and University of Colorado, Department of Physics, 440 UCB, Boulder, Colorado 80309, USA}

\date{\today}
	
	\begin{abstract}
		Thermodynamically induced length fluctuations of high-reflectivity mirror coatings put a fundamental limit on sensitivity and stability of precision optical interferometers like gravitational wave detectors and ultra-stable lasers. 
		The main contribution - Brownian thermal noise - is related to the mechanical loss of the coating material.
		  Al\textsubscript{0.92}Ga\textsubscript{0.08}As/GaAs crystalline mirror coatings are expected to reduce this limit.
		First measurements on cryogenic silicon cavities revealed the existence of additional noise contributions exceeding the expected Brownian thermal noise. We describe a novel, non-thermal, photo-induced effect in birefringence that is most likely related to the recently discovered birefringence noise. Our studies of the dynamics and power dependence are an important step toward uncovering the underlying mechanisms. 
        Averaging the anticorrelated birefringent noise results in a residual noise that is shown to be substantially different from Brownian thermal noise. 
        To this end, we have developed a new method for analyzing the coating noise in higher-order transverse cavity modes, which makes it possible for the first time to determine the contribution of Brownian thermal noise to the total cavity noise.
		The new noise contributions must be considered carefully in precision interferometry experiments using similar coatings based on semiconductor materials.
		\begin{description}
			\item[Subject areas]
			optical interferometry, photonics, semiconductor physics
		\end{description}
	\end{abstract}
		\maketitle

\section{\label{sec1}Introduction}

Optical interferometers are by far the most sensitive measuring devices: ranging from km-size gravitational wave detectors \cite{har06b,hil11,adh20,ET20} to cm-size ultra-stable resonators \cite{kes12a,hae15a,mat17a,zha17,oel19} for the best atomic clocks.
For all these applications, highly-reflective mirror coatings are essential. 
The fundamental displacement noise of the mirror surface, which leads to length fluctuations, must be minimized to reach the ultimate performance.
In the most sensitive frequency band of current gravitational wave detectors, coating noise is one of the main limitations on the strain sensitivity \cite{ET20}.
Optical coatings with lower noise level are indispensable for the tenfold enhanced sensitivity that the next generation gravitational wave detectors like the Einstein telescope \cite{ET20} are aiming for. 
Noise in optical coatings limits the linewidth of today's most frequency stable lasers to a few mHz at 1.5 $\mathrm{\mu m}$.
These lasers work as local oscillators for the most precise atomic clocks based on narrow-linewidth optical transitions.
Therefore, employing low noise optical coatings in ultra-stable lasers helps to exploit the potential of the $\mathrm{\mu Hz}$- or even $\mathrm{nHz}$-linewidth of atomic transitions \cite{oel19,lan21a,bot22}, which improves the clock stability for investigation of fundamental physics \cite{del17, ken20, lan21,deb20} and for a future redefinition of SI second \cite{rie15,lod19}.
Hence, a considerable number of studies have been carried out over the last decades to reduce mirror noise with novel coating materials \cite{tai20,pan18,ste18b,vaj18,mag18,rob21} and advanced mirror concepts \cite{kim08,dic18}.

According to the fluctuation-dissipation theorem \cite{cal51,kub66}, the Brownian thermal noise is related to the mechanical dissipation by internal friction \cite{lev98,ste18a} and it can thus be reduced by using coating materials with lower mechanical loss.
Despite significant efforts devoted to the development of improved optical coatings \cite{tai20,pan18,ste18b,vaj18,mag18,rob21} including doping \cite{har06} and annealing \cite{gra20a}, the mechanical loss coefficients $\phi$ of characterized conventional dielectric coatings so far have not been improved substantially \cite{top96}.

A promising approach for further reduction of Brownian thermal noise is provided by crystalline mirror coatings comprised of Al$_{0.92}$Ga$_{0.08}$As/GaAs multilayers.
These monocrystalline multilayers exhibit lower mechanical loss inferred from mechanical ring-down \cite{col12, pen19} than conventional dielectric coatings ($\phi \approx 4\times 10^{-4}$ \cite{yam06,rob21}).
In addition, the optical loss of these coatings has reached a comparable level to that of the dielectric coatings ($<$ 10 ppm), making them an attractive alternative to conventional dielectric coatings.
Crystalline coatings were expected to significantly improve the performance of ultra-stable lasers and gravitational wave detectors \cite{leg18,adh20}. 
So far, only few experimental data of their coating noise in optical interferometers \cite{col13} is available.
At room temperature, an ultra-stable optical resonator with these coatings has demonstrated lower noise than expected for a similar resonator with dielectric coatings \cite{col13}.
However, the large noise contribution from other cavity constituents (80\%) hindered accurate evaluation of the coating performance, determining the mechanical loss to $\phi = (4 \pm 4) \cdot 10^{-5}$. 

Studying the frequency stability of cryogenic silicon cavities employing these coatings we have recently discovered novel noise contributions exceeding the expected Brownian thermal noise  \cite{ked23}. 
While that paper has mostly concentrated on the dependence of the noise on laser power, here,
to gain more insight, we first investigate the photo-induced change in the birefringence and its possible contribution to noise. 
We find a novel non-thermal photo-birefringent effect, which is highly nonlinear in power and which shows a dynamic response with power-dependent timescales of up to several hours. Consequently we stabilize the power well enough, that the corresponding noise is negligible compared to the newly discovered noise sources.

Second, we investigate the spatial correlation of these noise sources by simultaneously probing different transverse cavity modes, thereby accessing coating noise independent of technical noise contribution. 
To this end we have developed a technique where two lasers from opposite sides of the cryogenic cavity are locked to different spatial and polarization eigenmodes of the cavity. 

Finally, analyzing the difference between the two HG-modes obtained by this method enables us for the first time to directly measure thermal noise of a Al\textsubscript{0.92}Ga\textsubscript{0.08}As/GaAs coating at 124 K, 
 and to demonstrate that the previously observed excess noise \cite{ked23} is not simply Brownian noise from unexpectedly large mechanical loss.

The measurements of the birefringent effects reported in this paper provide a critical lead for future investigations of the origin of the novel noise source presented in semiconductor materials.

\section{\label{sec2}Experiment}
The intrinsic birefringence of these crystalline coatings \cite{col13,win21} leads to a splitting of resonator polarization eigenmodes. Frequency noise associated with these individual modes needs special consideration. 
In our measurements we could separate three uncorrelated contributions to the fluctuations of the optical path length $d(t)$ for the two polarization eigenmodes averaged over the resonator mode area:

\begin{equation}
	d_\mathrm{slow/fast}(t) = d_\mathrm{Brown}(t) \pm d_\mathrm{birefr}(t) + d_\mathrm{global}(t).
	\label{eq:noise_source}
\end{equation}
It contains spatially uncorrelated Brownian noise $d_\mathrm{Brown}$, 
fluctuations of the coating birefringence $d_\mathrm{birefr}$ where $\pm$ applies to the fast and slow polarization eigenmode, and global excess noise $d_\mathrm{global}$ with a spatial correlation length larger than the beam diameter.  
As these contributions are temporally uncorrelated, the total power spectral density (PSD) of the optical length fluctuations $S_d$ is obtained as:

\begin{equation}
	S_d=S_{\mathrm{Brown}}+S_{\mathrm{birefr}}+S_{\mathrm{global}},
\end{equation}
which leads to the same $S_d$ for both polarization eigenmodes.

\begin{figure}[!ht]
	\centering\includegraphics[width=0.5\textwidth]{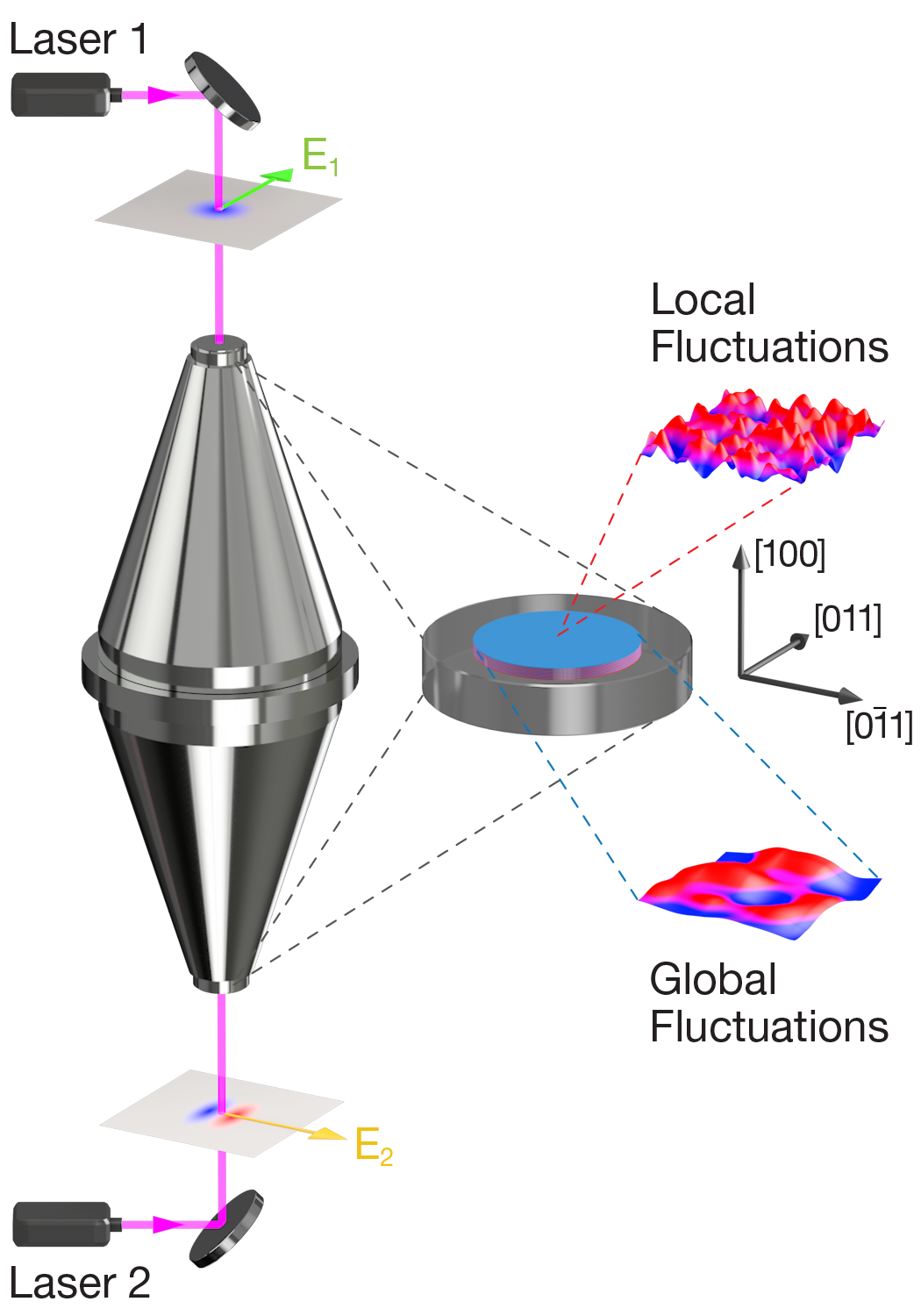}
	\caption{
		Schematic of experiment: two lasers are stabilized to the silicon resonator with crystalline mirror coatings. 
		The lasers can be stabilized to different polarization- and transverse cavity modes independently. 
		Local and global noise sources are depicted in arbitrary units. 
		The frequency fluctuations are measured by comparing the two lasers against a third laser stabilized to a similar reference cavity with dielectric coatings (Si2) \cite{mat17a}.}
	\label{fig:setup}
\end{figure}

The schematic overview of our experimental setup is illustrated in Fig. \ref{fig:setup}. 
Our optical resonators consist of mirrors with $\mathrm{Al_{0.92}Ga_{0.08}As/GaAs}$ crystalline coatings attached on a 21 cm long monocrystalline silicon spacer \cite{mat17a} operated at 124 K,  and on a 6 cm \cite{rob19} long spacer operated at 4 K or 16 K.
The first cavity utilizes two mirrors, both of which have a radius of curvature (ROC) of 2 m, resulting in a mode diameter of 964 $\mathrm{\mu m}$ on both mirrors. And the second cavity employs two concave mirrors, each with an ROC of 1 m, resulting in a smaller mode diameter of $588~\mathrm{\mu m}$.
Light propagates between the mirrors in vacuum, and any minute optical path length change $\Delta d$ can be precisely measured via the shift of the cavity resonance frequency $\Delta\nu$:

\begin{equation}
	\Delta d=-\frac{\Delta\nu}{\nu}\cdot L_\mathrm{cav},
\end{equation}
where $\nu=194$ THz is the laser frequency and  $L_\mathrm{cav}$ is the cavity length.

Thanks to the mechanical loss of single crystal silicon and the low coefficient of thermal expansion at our operating temperatures \cite{rob21}, the fundamental noise contributions from spacer and mirror substrates, including Brownian thermal noise \cite{kes12} and thermo-elastic noise \cite{bra99}, are one order of magnitude below the predicted coating Brownian noise (see Appendix \ref{sec:thermal_budget}), which makes these silicon resonators ideal platforms for investigating the coating performance.
As there are no measurements of the loss at 124 K, for the prediction we assume that the mechanical loss of these coatings at these cryogenic temperatures is nearly the same as at room temperature ($\phi \approx 2.5\times 10^{-5}$), even though there is a trend to lower loss at temperatures below 70 K \cite{col12}.  

Static birefringence has been observed in AlGaAs coatings. 
In our resonators, the [100] crystal direction of GaAs is normal to the mirror surface.
We observe that light polarized along the [011] crystal axis (slow axis) exhibits higher refractive index and propagates slower than light polarized along the [0$\bar{1}$1] crystal axis (fast axis), which is consistent with a recent report \cite{win21}.
For the 6 cm cavity, the mirrors were mounted with parallel orientation $\theta<3^{\circ}$ of the GaAs crystal axes, while for the 21 cm cavity a $\theta\approx15^{\circ}$ offset was later discovered. 
In the 21 cm cavity, this alignment of the coatings splits the resonances into two linearly polarized eigenmodes separated by $\Delta\nu_\mathrm{birefr}\approx200$ kHz, which is much larger than the cavity linewidth of $\Delta\nu_\mathrm{FWHM}=1.8$ kHz. 
The corresponding static birefringence of the coating multilayer $\Delta n_\mathrm{birefr} = n_\mathrm{slow}-n_\mathrm{fast}$ can be estimated as:
\begin{equation}
	\Delta n_\mathrm{birefr} = \frac{\Delta \nu_\mathrm{birefr}}{2 \nu}\cdot\frac{L_\mathrm{cav}}{ l_\mathrm{pen}}\cdot \frac{1}{\lvert\cos{\theta}\rvert},
	\label{eq:d_n}
\end{equation}
where $l_{\mathrm{pen}}=163$ nm is the penetration depth \cite{bab92} of the light field in the coatings, the factor 2 accounts for the two mirrors in the resonator,
and $\lvert\cos{\theta}\rvert$ is the correction factor for axis offset \cite{bra97a}.
The corresponding static birefringence ($690\pm3$ ppm) is similar to the 6 cm cavity ($731\pm 3$ ppm) and another room temperature optical resonator with crystalline coatings at $1.5\mathrm{\;\mu m}$ operated in our lab ($792\pm2$ ppm) and is slightly smaller than the value reported in \cite{col13} (1000 ppm).

We stabilize erbium doped fiber lasers (EDFL) to our cavities.
In the 21 cm cavity, two EDFLs are simultaneously locked on cavity resonances from both ends of the optical resonator via the Pound-Drever-Hall (PDH) technique \cite{bla01}. 
This enables us to investigate correlations in optical path length fluctuations between different polarization or transverse eigenmodes, which reveals spatiotemporal properties of the coating noise \cite{gra17}. 
The 6 cm cavity system is equipped with one laser only \cite{ked23}.

We systematically characterized and minimized all environmental and instrumental influences, typically referred to as ``technical noise'' (see Appendix \ref{sec:resonators}).
In total the technical noise contributions are reduced below the predicted Brownian thermal noise floor between 0.75 and 100 mHz Fourier frequency (see Appendix \ref{sec:tech_budget}).

\section{\label{sec3}Results}
\subsection{\label{subsec1}Photo-birefringent effect}
\begin{figure}[!ht]
	\centering\includegraphics[width=0.49\textwidth]{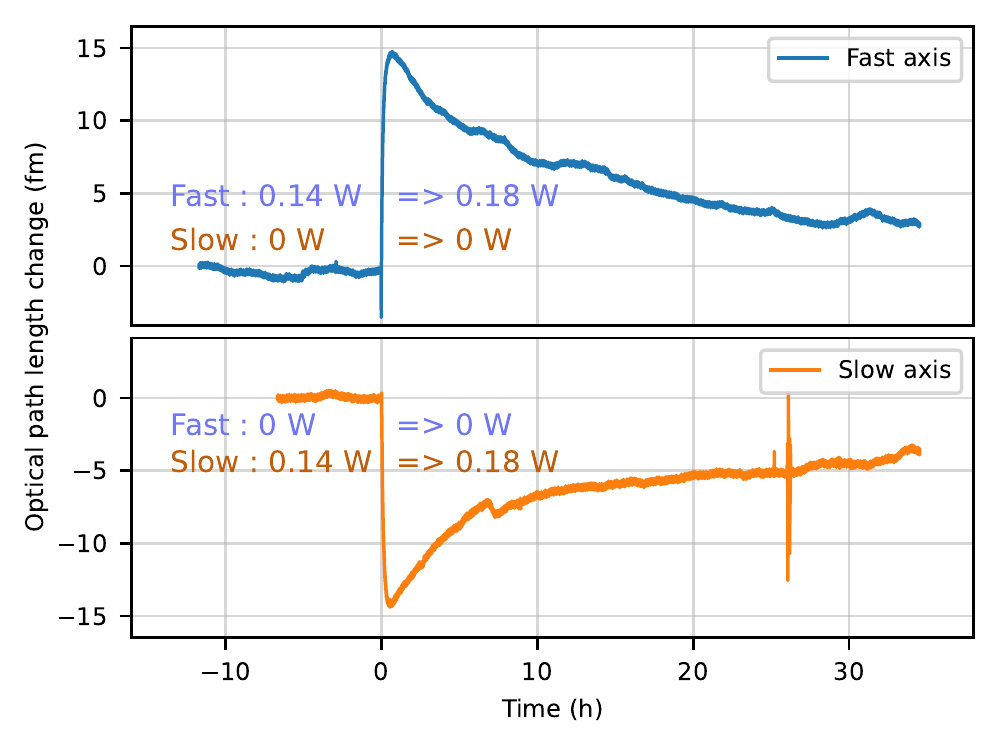}
	\caption{Single polarization transient response of the 21 cm cavity to a step change of intra-cavity power measured in succession. 
			The optical path length changes of the fast (blue) and the slow (orange) axes are symmetric.}
	\label{fig:transient_a}
\end{figure}

In dielectric \cite{far12} and crystalline \cite{cha16} mirror coatings, it was observed that intra-cavity power fluctuations lead to optical path length fluctuations.
To evaluate this effect in crystalline coatings at cryogenic temperature, we measured the frequency change in response to intra-cavity power. 
We observe that the response of optical path length to a step in the power is opposite for the two polarization eigenmodes (Fig. \ref{fig:transient_a}).
Initially the frequency quickly rises to a maximum with a time constant of a few hundred seconds, followed by a slower relaxation with a time constant of several hours. 
The sensitivity and time constant of the transient response strongly depend on the final intra-cavity power (see Appendix \ref{sec:PBR_power}). 
Due to the opposite sign between the two polarization axes, this photo-birefringent effect can be canceled with orthogonal alignment of the GaAs crystal axes. 

The photo-birefringent effect was not observed in previous studies on the photo-thermal response of these crystalline coatings \cite{cha16, her21}.
This is due to the fact that the photo-thermal effect in that study is more than 40 times stronger than the photo-birefringent effect observed in this work. 

The opposite sign of the transient response cannot be explained by a thermal effect from the absorbed laser power, which is the dominating process in dielectric coatings \cite{bra99,far12}, because temperature induced variation of optical path length is largely polarization independent.
We thus attribute this observation to a new light-induced change of birefringence in crystalline coatings (photo-birefringent effect).
A length change of $\Delta d=1\times 10^{-14}$ m in Fig. 2 corresponds to a change of birefringence $\delta n_\mathrm{birefr} = \delta n_\mathrm{slow}-\delta n_\mathrm{fast}=3\times 10^{-8}$ according to Eq. \ref{eq:d_n} (45 ppm of the static birefringence $\Delta n_\mathrm{birefr}$).

Such behavior is not seen in our otherwise identical reference silicon resonators with dielectric coatings (Si2 and Si4) \cite{mat17a,rob19}, therefore, this photo-birefringent effect results presumably from the semiconducting properties of the crystalline coatings.

While a full theoretical model has not yet been developed, we speculate that the photo-birefringent effect may be related to the linear electro-optic effect \cite{nam61}, as light-induced birefringence has been observed in other materials \cite{twu18,min07,agu06}.

For crystalline coatings, the crystal orientation of the photo-birefringent effect can be explained by a change in the electric field of 3 kV/m perpendicular to the coating surface, which is a relatively small magnitude compared to the 100 times larger electric field strength observed in some heterojunctions \cite{est02}.
\begin{figure}[!ht]
	\centering\includegraphics[width=0.5\textwidth]{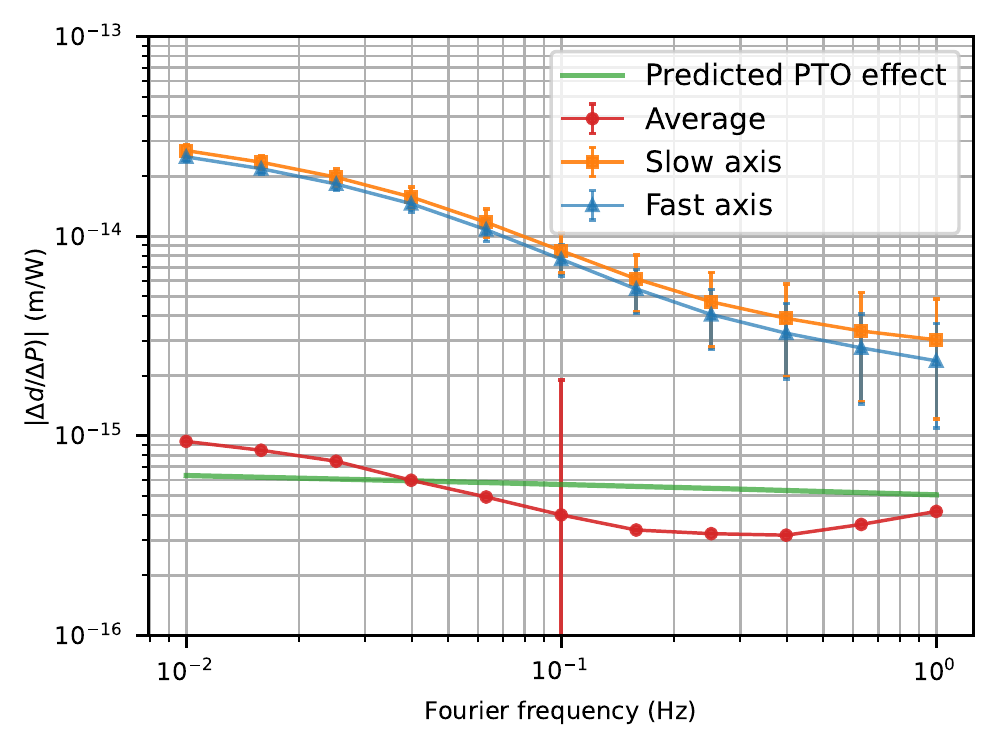}
	\caption{
    Frequency dependence of optical path length changes to intra-cavity power variations for fast (blue) and slow (orange) axes and for their average (red) at mean intra-cavity power of 0.54 W in comparison to the theoretical photo-thermal optic response (green), calculated according to \cite{far12}. 
    The error bars indicate the 95\% confidence interval.
    It includes statistical uncertainty and possible contributions due to slowly varying electronic offsets in the measurement. 
    The latter contribution is estimated by forwarding typical electronic offsets to the slope of PDH error signal.  
    For the red curve only one representative error bar is shown.
    }
	\label{fig:transient_b}
\end{figure}

To investigate the power dependence of the photo-birefringent effect, we have changed the intra-cavity power of the laser locked on slow axis in the three steps ($\mathrm{0.6~\Rightarrow~0.2~W}$, $\mathrm{0.6~\Rightarrow~1.6~W}$ and $\mathrm{1.6~\Rightarrow~0.6~W}$), and no optical power is coupled to the fast axis.
\begin{figure}[!ht]
	\centering\includegraphics[width=0.45\textwidth]{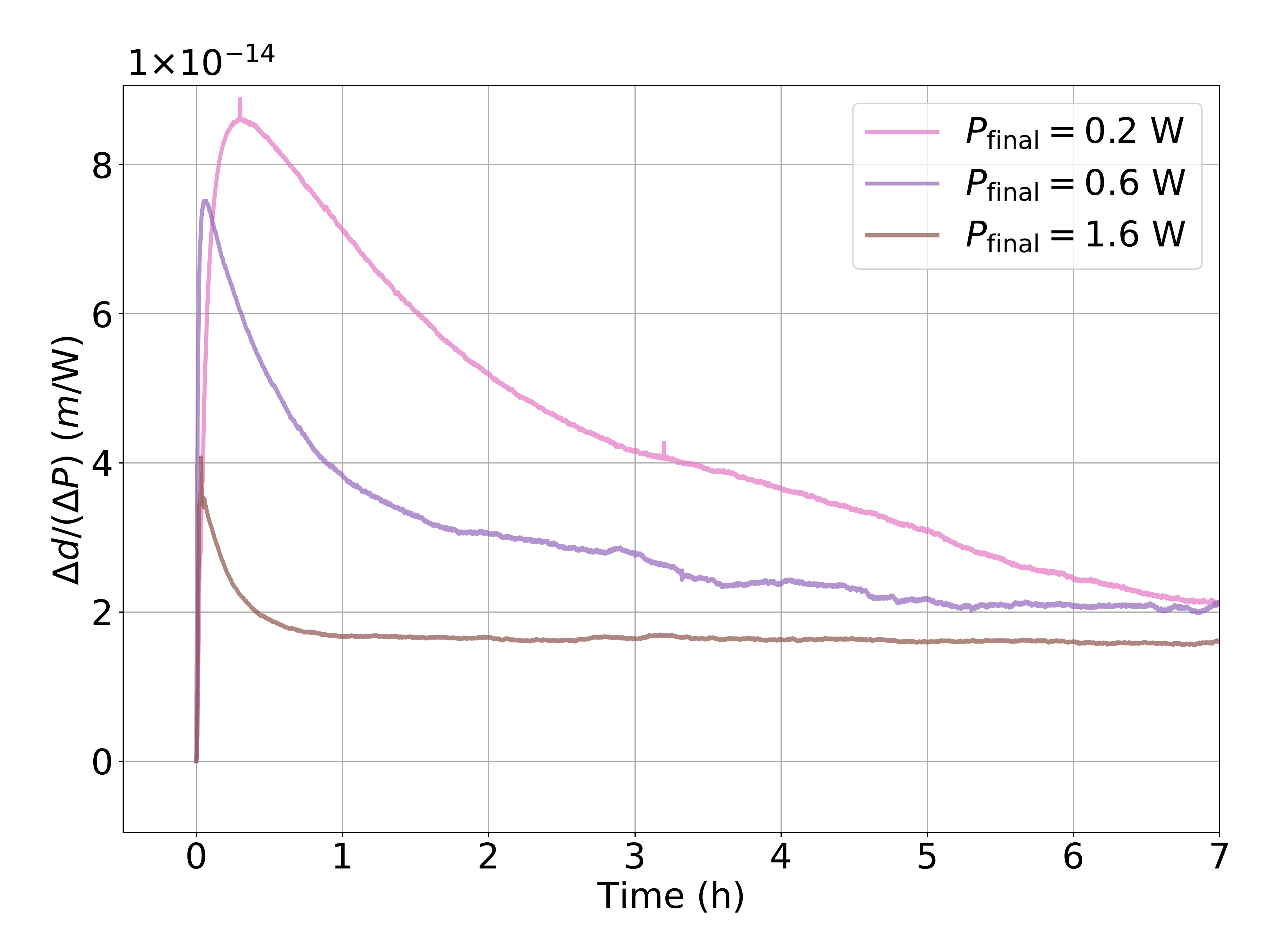}
	\caption{Normalized response of the optical path length $\Delta d$ (slow axis) to a step $\Delta P$ in intra-cavity optical power. Both amplitude and time constant show a strong dependence on the final intra-cavity power $P_\mathrm{final}$. }
	\label{fig:PB_a}
\end{figure}
The path length change is inferred from the observed change in optical frequency $\Delta \nu $ as $\Delta d = - L_\mathrm{cav} \Delta \nu / \nu$.
The normalized transient response of the optical path length shown in Fig. \ref{fig:PB_a} indicates a strong dependence of its amplitude and time constant on the final optical power.

To evaluate the influence of laser power noise on the optical path length fluctuations imposed by the photo-birefringent effect, we measured the small-signal transfer function from power to frequency (Fig. \ref{fig:transient_b}).
The measured transfer functions for two polarization axes are very similar as they are dominated by the photo-birefringent effect. 
The average of their complex amplitudes is the transfer function of the polarization averaged response after removing the photo-birefringent effect. 
This transfer function is compatible with that of the photo-thermal optical effect \cite{far12}. 
Due to the large uncertainty in the small difference, the remaining frequency dependency is insignificant.
For the two polarization eigenmodes, the sensitivity of the optical path length to power fluctuations decreases with increasing intra-cavity power (see Appendix \ref{sec:PBR_power}).

The decrease in the transfer function towards high frequencies corresponds to the fast initial step response (Fig. \ref{fig:transient_a}). 
The long-term behavior is not visible in the single-polarization transfer functions due to  limitations of the lowest measurable frequencies.

From these transfer functions we conclude that with our actively stabilized optical power, the optical path length noise related to the photo-birefringent effect of 
$S_d= 2\times10^{-36}~\mathrm{m^2/Hz} \cdot (f/\mathrm{Hz})^{-1}$ 
is suppressed well below the predicted coating Brownian thermal noise (see Appendix \ref{sec:tech_budget}).

\subsection{\label{subsec2}Birefringent noise}

\begin{figure*}[!ht]
	\centering\includegraphics[width=\textwidth]{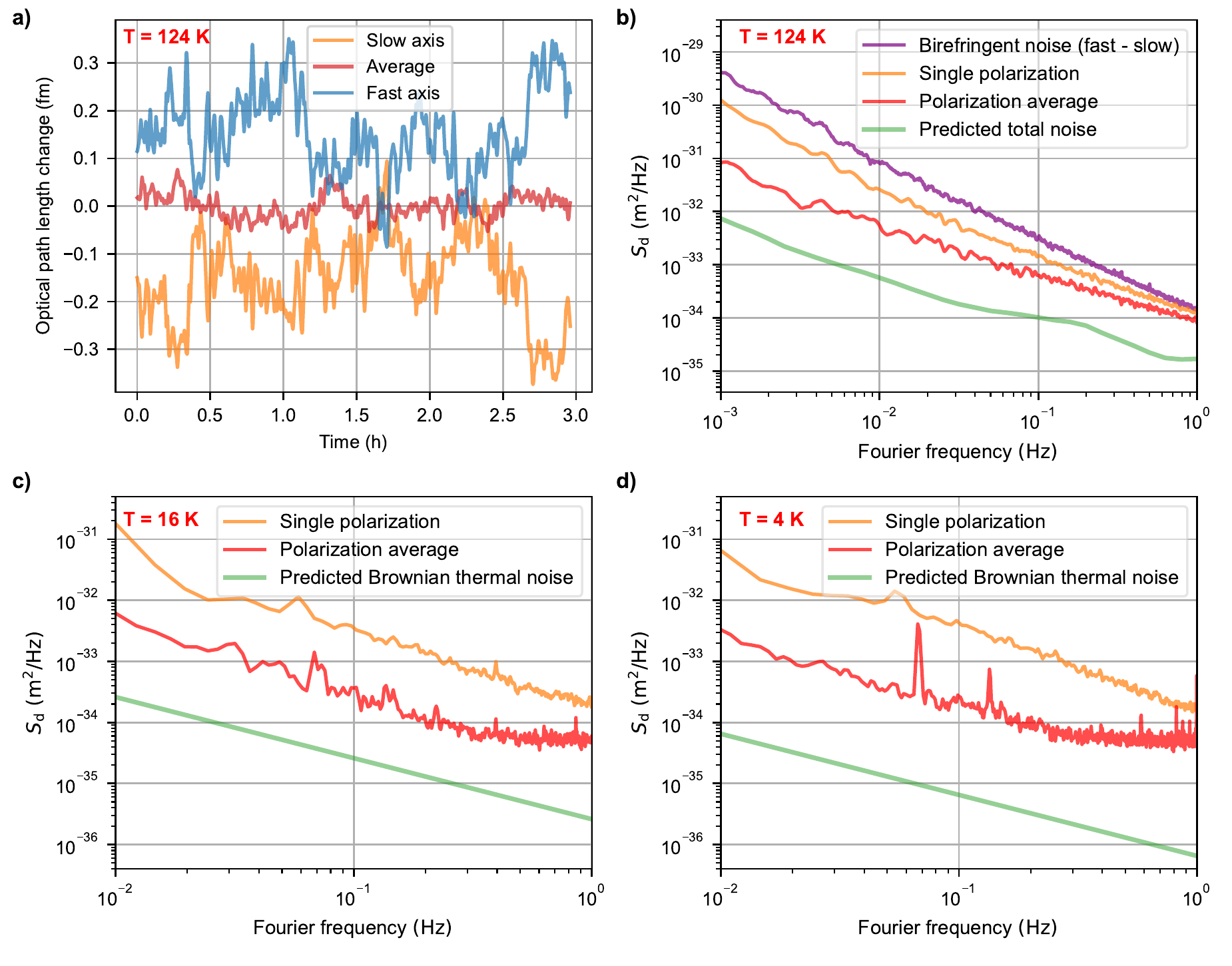}
	\caption{
		(a) Optical path length fluctuations of the 21 cm cavity measured with the beat signal between Si2 and the two polarization eigenmodes with intra-cavity power of 1.3 W in the fast and 0.7 W in the slow axis (blue, orange). 
		The noise from Si2 is below the  average of the two polarization eigenmodes (red). 
		(b) Power spectral densities $S_d$ of the length fluctuations in the 21 cm cavity. 
		Birefringent noise (purple), noise of an individual polarization eigenmode (orange) and average of two polarizations (red). 
		The sum of technical noise contribution and the predicted Brownian thermal noise for the Hermite-Gaussian HG\textsubscript{00}-mode (green) is included for comparison. 
		(c) and (d) Power spectral densities $S_d$ of the length fluctuations in the 6 cm cavity at 16 K (c) and 4 K (d). 
		Frequency stability of individual polarization eigenmodes (orange), of the average of two polarization eigenmodes (red) and the predicted Brownian thermal noise (green). The contribution of reference lasers has been removed from the PSDs (see Appendix \ref{A1}). }
	\label{fig:birefr}
\end{figure*}

With all relevant technical noise sources suppressed below the expected Brownian thermal noise, we lock the two independent lasers to the fundamental Hermite-Gaussian (HG) mode of the 21 cm cavity with different polarizations. 
The mirror coating noise is observed as the frequency fluctuations of the two lasers compared to the Si2 reference laser (Fig. \ref{fig:birefr}a). 

We observe strongly anti-correlated frequency fluctuations between the two polarization eigenmodes.  
Observing the frequency difference of the two lasers suppresses all common mode noise contributions leaving only the birefringence fluctuations (purple curve in Fig. \ref{fig:birefr}b). 

These so far unobserved intrinsic birefringence fluctuations (``birefringent noise'') in crystalline coatings lead to optical path length fluctuations with PSD two orders of magnitude higher than the predicted Brownian thermal noise (Fig. \ref{fig:birefr}b).
The same behavior is observed in the 6 cm cavity at 16 K and 4 K (Fig. \ref{fig:birefr}c-d). 
In Fig. 5b-d, the impact of polarization averaging gradually diminishes at higher Fourier frequencies as other noise contributions come into play. 
For the 124 K measurement, a common mode noise (see section II.C) becomes a significant contribution to the total coating noise. 
While in Fig. 5c and d, the limit originates from a technical noise as discussed in Ref \cite{ked23}.
We also find that the birefringent noise slightly increases with optical power \cite{ked23}.

Unlike photo-birefringent noise, the birefringent noise will not be suppressed by orthogonal alignment of the GaAs crystal axes, because the birefringence fluctuations of individual crystalline coatings are uncorrelated.

\subsection{\label{subsec3}Polarization-independent noise contributions}

By simultaneously probing the two polarization eigenmodes with two independent lasers, we are able to remove the anticorrelated birefringent noise by polarization averaging.
Nevertheless, we observe at all three temperatures a remaining noise level that is still significantly higher than the predicted coating Brownian thermal noise.
This remaining noise is independent of optical power, which indicates a noise mechanism different from birefringent noise described in the previous section. 
\begin{figure*}[!ht]
	\centering\includegraphics[width=\textwidth]{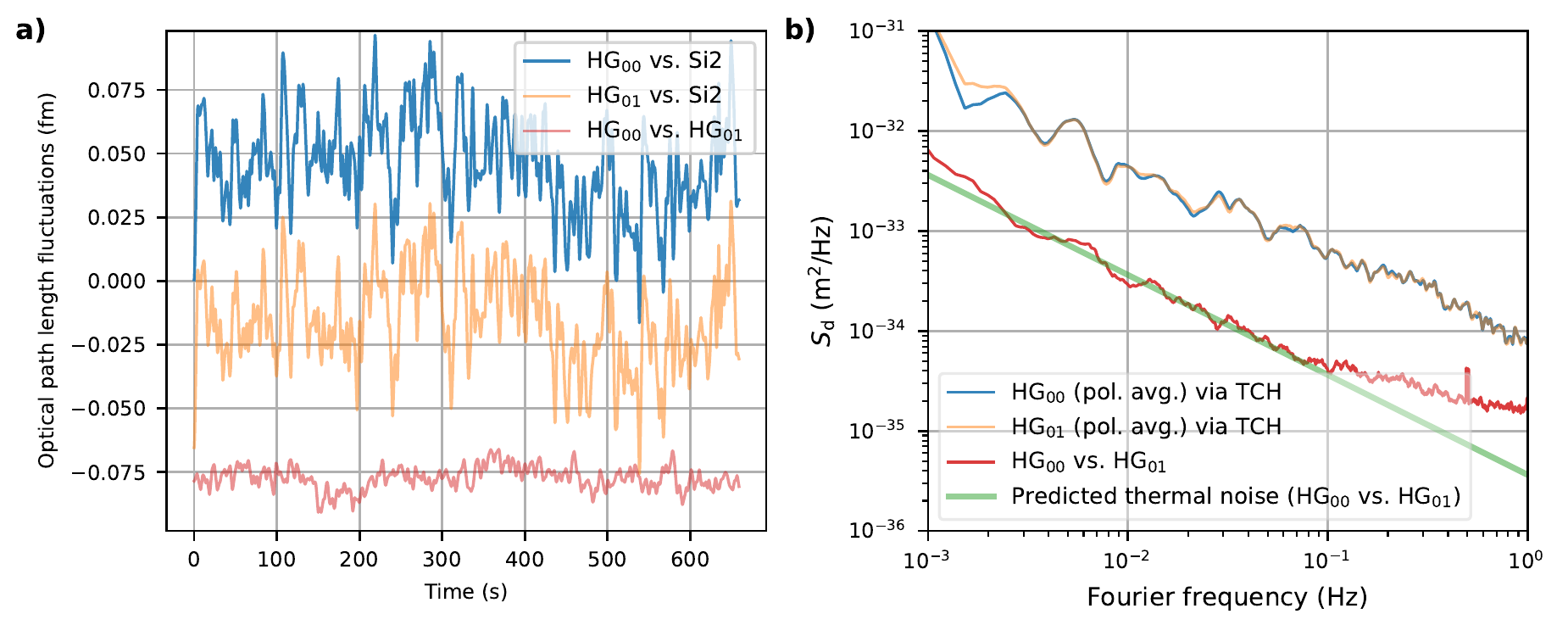}
	\caption{
		(a) Measured optical path length fluctuations of polarization averaged HG\textsubscript{00} (blue) and HG\textsubscript{01} mode (orange) referenced to another cavity (Si2), and the difference between the two HG-modes (red). For clarity the curves have been shifted by an arbitrary amount. 
		(b) Spectral power densities of the individual displacement noise of HG\textsubscript{00} (blue) and HG\textsubscript{01} mode (orange) determined by three-cornered-hat (TCH) method, and of their difference (red). The estimated differential Brownian noise between the two modes is shown in green.
	}
	\label{fig:pol_indep}
\end{figure*}
The remaining noise roughly has a $1/f$ dependency in PSD akin to Brownian thermal noise (Fig. \ref{fig:birefr}, red lines). 
This noise level could be in principle explained by an increased coating mechanical loss $\phi$. 
While there are reliable data for the mechanical loss from cantilever ring-down measurement at high frequencies at 4 K, 16 K, and 300 K, no direct loss measurements of coatings at low frequencies and at 124 K are available.
To measure the relevant loss $\phi_\mathrm{124K}$, we analyze the spatial correlations of the polarization independent remaining noise.

For dielectric coatings, Brownian thermal noise is the leading spatial uncorrelated noise source (local noise) with a correlation length on the order of coating thickness, and thus it shows up in the difference in displacement fluctuations between two different HG modes.
Global noise sources with correlation length much larger than the mode diameter of 1 mm are common to both modes and are strongly suppressed in this difference \cite{gra18}.

The technical difficulty associated with this method in crystalline coatings is that the birefringent noise is also local noise (see Appendix \ref{sec:BR_spatial}):
to investigate the spatial correlation of polarization averaged remaining noise between HG\textsubscript{00}- and HG\textsubscript{01}-mode, four lasers would be required to average the two polarization eigenmodes of both HG-modes respectively, resulting in considerable additional complexity.

To solve this problem, we developed a dual-frequency locking technique that enables the cancellation of birefringent noise using only one laser for each HG-mode.
This is achieved by simultaneous excitation of both polarization eigenmodes using additional spectral lines generated with an electro-optic modulator.
In this way, an overall error signal containing equally weighted contributions from both polarization eigenmodes is generated, thus the laser can be stabilized to their average.
With this dual-frequency locking technique, we suppress the birefringence noise by more than two orders of magnitude, and the locking noise is well below the predicted coating Brownian thermal noise.
More details about this technique can be found in Appendix \ref{sec:2FreqLock}.

We stabilize one laser coupled from the top of the cavity to the polarization averaged HG\textsubscript{00} mode.
The other laser is stabilized simultaneously to the averaged HG\textsubscript{01} mode from the bottom of the cavity. 
Even though there is a certain overlap of these two modes, they probe fluctuations averaged over significantly different areas of the mirror coatings.  
The displacement fluctuations are measured by referencing the two laser frequencies to the Si2 system (Fig. \ref{fig:pol_indep}a blue and orange). 
These measured fluctuations contain (similar) contributions from the cavity with crystalline coatings and from the Si2 reference cavity (see Appendix \ref{sec:BTN_spatial}). 

The fluctuations visible in the direct difference between the two HG-modes (Fig. \ref{fig:pol_indep}a red) contain only local noise such as Brownian thermal noise, but no correlated global noise (Eq. \ref{eq:noise_source}).  
Thus, this difference constitutes an upper bound on the coating Brownian thermal noise. 
Fig. \ref{fig:pol_indep}b compares the power spectral densities $S_d^{(\Delta)}$ of the measured fluctuations against that predicted from coating Brownian thermal noise.  
The measured displacement noise in the difference of the two modes (Fig. \ref{fig:pol_indep}b red) corresponds to a loss coefficient $\phi_\mathrm{300K}\approx 2.5\times 10^{-5}$.
Loss coefficients from ringdown measurements are only available near room temperature ($\phi_\mathrm{300K}\approx 2.5\times 10^{-5}$ \cite{col13}, $\phi_\mathrm{300K}\approx 4.78 (5)\times 10^{-5}$ \cite{pen19}) and below 70 K \cite{col12} with a trend towards lower loss at low temperatures.
Thus, the mechanical loss relevant for precision interferometry at 124 K shows no unexpected behavior  to the value obtained from mechanical ring-down at different frequencies and temperatures. 

With this precisely determined local noise $S_d^{(\Delta)}$ we can calculate its contribution to the individual displacement fluctuations averaged over the HG\textsubscript{00} and HG\textsubscript{01} modes.
The corresponding PSDs are $S_\mathrm{local}^{(00)}=1.33\times S_d^{(\Delta)}$ for HG$_{00}$ mode and $S_\mathrm{local}^{(01)}=1.00\times S_d^{(\Delta)}$ for HG$_{01}$ mode (see Appendix \ref{sec:BTN_spatial}).
The PSD in Fig. \ref{fig:pol_indep}b obtained from three-cornered-hat method, where contribution from Si2 is removed, clearly shows that the observed noise in the individual modes is one order of magnitude larger than the value calculated from these numbers ($1.33\times S_d^{(\Delta)}$).

Therefore, the displacement noise experienced by the modes is dominated by a non-local noise process with spatial correlation lengths larger than the mode size. 
This global excess noise $d_\mathrm{global}$ also appears to be a persistent source of noise at 4 K and 16 K, as recorded also for the 6 cm cavity (Fig. \ref{fig:birefr}b and c) \cite{ked23}. 

In principle technical noise or hitherto unobserved noise, e.g. from optical contacts, 
could appear as such a global noise. 
However, all known technical noise contributions have been found to be significantly lower than the observed global noise (see Appendix \ref{sec:tech_budget}).
Moreover, the detection of this novel global noise in two independent systems in separate laboratories makes underestimated technical noise unlikely.
Additionally, the close agreement between the experimentally observed noise in all our cavities using conventional coatings \cite{mat17a,rob19} and the theoretically expected noise based on loss measurements of these dielectric coatings leaves little room for additional noise sources of comparable magnitude. 

In summary, these facts make us confident that this novel global excess noise is most likely intrinsic to the crystalline coating.

\section{Conclusion and discussion}\label{sec4}

We have investigated the properties related to the recently discovered noise in AlGaAs crystalline mirror coatings \cite{ked23} under conditions relevant for precision interferometry with two different cryogenic silicon cavities at 4~K, 16~K and 124~K. 

We discovered a non-thermal photo-birefringent effect, which is a change in the static coating birefringence depending on laser power at the mirror. 
We could show that with sufficiently stabilized laser power its contribution to the interferometer noise is negligible and thus this effect can not explain the recently discovered noise in AlGaAs coatings.

We can distinguish three fundamental contributions to the novel noise.
The biggest contribution is birefringent noise. 
We have presented and evaluated a technique to cancel this noise by averaging both polarization eigenmodes.

We have investigated the $1/f$ global excess noise that remains after cancelling birefringent noise. It was characterized by comparing different spatial modes and we have shown that it is not related to technical noise. 
Most likely it is associated with the semiconductor properties of the coatings.

With this method we unambiguously show that the coating Brownian noise at 124~K is in very good agreement with the theoretical prediction obtained from the room temperature mechanical loss factor, thus confirming the expected Brownian noise reduction in AlGaAs crystalline coatings.

The first two contributions are significantly higher than the Brownian thermal noise which has significant implications on the use of current crystalline coatings in future ultra-sensitive interferometers.

In our ultra-stable lasers based on cryogenic silicon cavities, crystalline coatings suffer from birefringent noise and therefore exhibit significantly inferior performance than dielectric coatings.
After suppressing the birefringent noise by polarization-averaging, the PSD of crystalline coatings is reduced compared to conventional dielectric coatings, with a ratio of 
$S^{(\mathrm{crys})}_d / S^{(\mathrm{dielec})}_d = 0.77$ at 4 K, 
0.38 at 16 K and 0.83 at 124 K.
The large improvement at 16 K stems from two factors: 1) the global excess noise does not show strong temperature-dependence while thermal noise increases linearly with temperature. 2) The smaller mode area in the 6 cm cavity would lead to pronounced Brownian thermal noise if dielectric coatings are utilized.
For cavities operating at room temperature or other wavelengths, further investigations are still required.

Third generation cryogenic gravitational wave detectors such as the low frequency Einstein Telescope (ET-LF) \cite{ET20} are proposed to operate at the quite similar temperature, wavelength and intra-cavity laser intensity as the 6 cm cavity, except the 300 times larger beam radius.
Extrapolating our results, we conclude that the current crystalline coatings would lead to higher noise than the dielectric coatings in these systems (see Appendix \ref{sec:ET}), even if the birefringent noise could be canceled and the correlation length of the global excess noise would be only on the order of 1 mm - the lower limit deduced from our measurement - which would allow to spatially average the excess noise.
Similarly, by extrapolating our results from our 21 cm resonator, we have estimated the potential performance of crystalline coatings in LIGO Voyager \cite{adh20}.
Our findings indicate that current crystalline coatings would result in approximately 90\% higher noise PSD than conventional dielectric coatings.
Therefore, current crystalline coatings have no advantage over the conventional dielectric coatings in third generation cryogenic gravitational wave detectors.

As the mechanisms for the two new noise processes might be related to defects and impurities \cite{har08,bor99} of the semiconductor coating, a better understanding of the microscopic effects could lead to a reduction of the noise.
This knowledge will also be helpful to other semiconductor-based coatings, such as aSi/SiN \cite{ste21}, that are currently discussed for precision interferometry. 
Our findings indicate that a broader investigation of noise processes in a wider class of semiconductor materials is important which might shed further light on the underlying mechanisms.

\begin{acknowledgments}
	We acknowledge support by the Project 20FUN08 NEXTLASERS, which has received funding from the EMPIR programme cofinanced by the Participating States and from the European Union’s Horizon 2020 Research and Innovation Programme, 
	and by the Deutsche Forschungsgemeinschaft (DFG, German Research Foundation) under Germany’s Excellence Strategy–EXC-2123 QuantumFrontiers, Project-ID 390837967; SFB 1227 DQ-mat, Project-ID 274200144. 
	This work is partially supported by the Max Planck-RIKEN-PTB Center for Time, Constants and Fundamental Symmetries.
	This work is also supported by NIST, DARPA, AFRL, JILA Physics Frontier Center (NSF PHY-1734006).
\end{acknowledgments}%

\appendix

\section{Experimental setups}\label{A1}
\subsection{Cryogenic silicon resonators}\label{sec:resonators}

The 21 cm silicon resonator (Si5) in this experiment was set up and operated at PTB, Germany to investigate the noise from crystalline mirror coatings. 
The mirror pair reaches finesse values of $(3.65\pm0.01)\times10^{5}$ for the fast axis and $(3.58\pm0.01)\times10^{5}$ for the slow axis, corresponding to a total loss of 17.4 ppm and 17.8 ppm respectively.
The Si5 setup is based on our previous design of silicon resonators that are equipped with dielectric coatings (Si2 and Si3).
The temperature of Si5 is also controlled with cold nitrogen gas \cite{hag13a}.
Si2 and Si3 demonstrated the Brownian thermal noise limited performance of 
$S_d= 8\times10^{-35}~\mathrm{m^2/Hz} \cdot (f/\mathrm{Hz})^{-1}$ 
between 0.01 and 10 Hz Fourier frequency after suppression of technical noise \cite{mat17a} (Si3 is later transferred to our lab in JILA).\\
Compared to previous systems, we further suppressed technical noise in Si5:
residual amplitude modulation (RAM) is suppressed in both laser systems from both ends \cite{zha14}.
Optical path length fluctuations in the setup (fiber and free space) are actively canceled \cite{ma94} besides a free space path of less than 30 cm.
This ensures much lower fiber noise level compared to the predicted Brownian thermal noise. 
Frequency fluctuations of the two lasers are measured by referencing them to Si2 \cite{mat17a}. 
Frequency fluctuations between the two lasers are detected directly by a photodetector, where the transmitted light of the far end laser and the reflected sidebands of the near end laser interfere, thus avoiding any uncompensated optical path. 
We actively suppress the seismic noise with a commercial antivibration platform, and improve it for Fourier frequencies above 0.1 Hz with an additional low-frequency feedback loop including high performance seismometer and tiltsensor \cite{kaw21}. 
Parasitic etalons are identified by correlating ambient pressure with frequency fluctuation of the resonator at Fourier frequencies below 10 mHz, and are reduced by tilting optical surfaces and adding optical isolators.
With technical noise suppressed, Si5 facilitates the investigation on the Brownian thermal noise from the two crystalline coatings of 
$S_d= 1\times10^{-36}~\mathrm{m^2/Hz} \cdot (f/\mathrm{Hz})^{-1}$ 
(see Appendix \ref{sec:tech_budget}). 

To determine the PSDs of frequency noise from Si5 in Fig. \ref{fig:birefr}b and Fig. \ref{fig:pol_indep}b, we apply the three-cornered-hat analysis \cite{pre93} to the PSD.
This method requires two additional independent reference resonators, involving Si2 and a second reference laser at 698 nm stabilized to a 48 cm room-temperature cavity with dielectric coatings attached on spacer made of ultra-low-expansion glass \cite{hae15a}, and the frequency gap is bridged by an optical frequency comb.
We measure beatnotes between these systems (Si5-Si2, Si5-ULE, and Si2-ULE) with lambda-type zero-dead-time frequency counters, and calculate PSDs of their frequency fluctuations.
Assuming uncorrelated noise in the three systems, the noise PSD of Si5 can be obtained as: 
\begin{equation}
S_d^\mathrm{(Si5)}=\frac{1}{2}\left(S_d^\mathrm{(Si5-Si2)}+S_d^\mathrm{(Si5-ULE)}-S_d^\mathrm{(Si2-ULE)}\right).
\end{equation}

The 6 cm resonator with crystalline coatings (Si6) is located in JILA, USA. 
This system is based on the design of Si4 \cite{rob19}, which is cooled with a closed cycle cryostat, and its temperature can be varied between 4 K and 16 K. 
The performance of Si6 is determined by subtracting the noise of Si3, which is carefully characterized with a strontium lattice clock in the same lab \cite{oel19}, from the beat with Si3.

Detailed characterization of various technical noise is reported in Ref. \cite{ked23}. 
Briefly, the 6 cm cavity has a birefringent mode splitting of 770 kHz. An interrogation scheme different from Si5 is employed, where both polarization modes are simultaneously probed with phase modulation sidebands of a single laser to cancel the birefringent noise. 
The residual noise is found to be of similar scale to that of Si5.  
Comparison between Si5 (124 K) and Si6 (4.7 K and 16.7 K) thus offers a unifying view of the residual global noise.  
Si6 realizes a frequency stability of $\mathrm{mod} ~\sigma_y=5.5\times 10^{-17}$ with excellent long term performance.

A brief summary of the cryogenic silicon resonators is given in Table \ref{table:cavities}:
\begin{table}[ht!]
	\caption{Key differences between cryogenic silicon cavities.}
	\label{table:cavities}
	\begin{ruledtabular}
		\begin{tabular}{c|cccc}
			Name & Si1/2/3 & Si4 & Si5	&Si6	\\
			\hline
			Cavity lentgh (m) & 0.21 & 0.06 & 0.21 & 0.06 \\
			Temperature (K) & 124 & 4/16 & 124 & 4/16 \\
			Optical coating & Dielec. & Dielec & Crys. & Crys. \\
		\end{tabular}
	\end{ruledtabular}
\end{table}

\subsection{Thermal noise budget of the 124~K system}\label{sec:tech_budget}

\begin{figure}[ht!]
	\centering\includegraphics[width=0.45\textwidth]{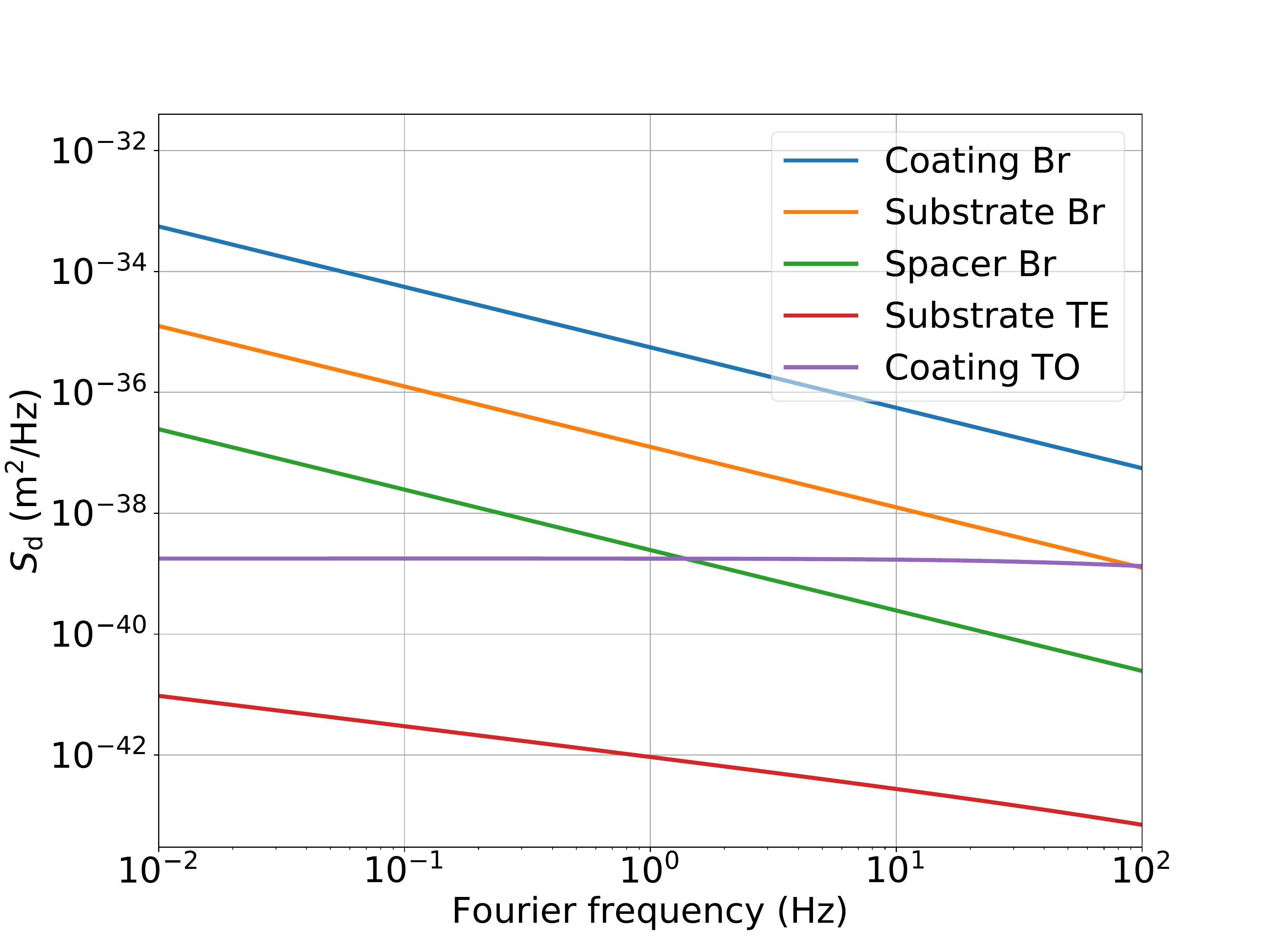}
	\caption{Thermal noise contributions in the Si5 resonator, Br: Brownian thermal noise, TE: thermo-elastic noise and TO: thermo-optic noise.}
	\label{fig:CBN}
\end{figure}

The thermal noise budget for the power spectral density of length fluctuations $S_d(f)$ in the 21 cm resonator (Si5) with AlGaAs mirror coatings is illustrated in Fig. \ref{fig:CBN}.
The Brownian thermal noise of the cavity constituents are calculated with the equations from ref. \cite{kes12} while the material properties are taken from \cite{hag13a,col13}.
Thermo-optic noise is calculated by averaging the thermal expansion and refractive index change, which are induced by the thermal fluctuations, across the mode area \cite{eva08}.
The Brownian thermal noise contributions from the AlGaAs coating is by far the biggest contribution for all considered frequencies. 

\subsection{Resonator parameters}\label{sec:parameters}
The resonator parameters used for the calculation of thermal noise and photo-optic response are summarized in the following tables. Table \ref{table:Si5_para} for properties of the optical resonators and Table \ref{table:c_para} for coating parameters.
\begin{table}
		\caption{Parameters for optical resonators.}
		\label{table:Si5_para}
\begin{ruledtabular}
    {\centering
    \begin{tabular}{c|c}
        Parameter &  value \\
    \hline
        \multicolumn{2}{c}{21 cm cavity} \\
    \hline
        Cavity length & 0.212 m  \\
    \hline
        Spacer radius & 0.04 m  \\
    \hline
        Radius of central bore  & 5 mm  \\
    \hline
        ROC of mirror & 2 m  \\
    \hline
        Beam radius on mirror & 482 $\mathrm{\mu m}$  \\
    \hline
        Cavity temperature & 124 K  \\
    \hline			
        Cavity finesse & $3.6\times10^{5}$  \\
    \hline			
        Laser wavelength & 1542 nm  \\
    \hline	
\hline
    \multicolumn{2}{c}{6 cm cavity} \\
    \hline
        Cavity length & 0.06 m  \\\hline
        ROC of mirror & 1 m  \\\hline
        Beam radius on mirror & 294 $\mathrm{\mu m}$  \\\hline
        Cavity temperature & 4 K or 16 K \\\hline			
        Cavity finesse & $2.9\times10^{5}$  \\\hline			
        Laser wavelength & 1542 nm  \\\hline	
    \hline
    \multicolumn{2}{c}{Single crystal silicon} \\\hline
        Young's modulus & 188 GPa \cite{bra73}  \\\hline	
        Poisson ratio & 0.26 \cite{bra73}  \\\hline
        Density & $2331~\mathrm{kg/m^3}$ \cite{mcs53}   \\\hline
        Thermal conductivity  & 600 $\mathrm{W/m\cdot K}$ \cite{gla64}  \\\hline
        Specific heat & $330~\mathrm{J/kg\cdot K}$   \cite{flu59}\\\hline
        Mechanical loss & $0.83\times10^{-8}$ \cite{naw08}  \\
\end{tabular}
}
\end{ruledtabular}
\end{table}

\begin{table}
		\caption{Parameters for the crystalline coating.}
		\label{table:c_para}
       \begin{ruledtabular}
 
{\centering
    \begin{tabular}{c|c}
Parameter &  value \\
\hline
\multicolumn{2}{c}{Optical coating} \\\hline
Coating structure & GaAs + 45$\cdot$(AlGaAs+GaAs)\\\hline
Layer optical length & quarter wavelength\\\hline
coating total thickness & $11.68~\mathrm{\mu m}$  \\\hline
coating mechanical loss & $2.5\times10^{-5}$\cite{col13,pen19}  \\\hline	
\multicolumn{2}{c}{Al$_{0.92}$Ga$_{0.08}$As} \\\hline
Layer thickness & 132.634 nm  \\\hline
Young's modulus & 83 GPa \cite{ada93}  \\\hline	
Poisson ratio & 0.40 \cite{ada93}  \\\hline
Density& $3885~\mathrm{kg/m^3}$ \cite{ada93}   \\\hline
Refractive index & 2.9065 \cite{jen90}  \\\hline
Temperature coefficient of n & $0.99\times10^{-4}$/K \cite{jen90}  \\\hline
Thermal conductivity & 69 $\mathrm{W/m\cdot K}$ \cite{afr73}  \\\hline
CTE (GaAs) & $3\times10^{-6}$/K   \\\hline
Specific heat & $313~\mathrm{J/kg\cdot K} $   \\\hline
\hline
\multicolumn{2}{c}{GaAs} \\\hline
Layer thickness & 115.477 nm  \\\hline
Young's modulus & 86 GPa \cite{ada93}  \\\hline	
Poisson ratio & 0.31 \cite{ada93}  \\\hline
Density & $5317~\mathrm{kg/m^3}$ \cite{ada93}   \\\hline
Refractive index & 3.3383 \cite{bla82c}  \\\hline
Temperature coefficient of n & $1.75\times10^{-4}$/K \cite{bla82c}  \\\hline
Thermal conductivity & 100 $\mathrm{W/m\cdot K}$ \cite{bla82c}  \\\hline
CTE & $3\times10^{-6}$/K \cite{ada93}   \\\hline
Specific heat& $215~\mathrm{J/kg\cdot K}$ \cite{bla82c}  \\
\end{tabular}
}
  \end{ruledtabular}
 
\end{table}

\subsection{Technical noise budget of the 124~K system}
\label{sec:thermal_budget}
\begin{figure}[!ht]
	\centering\includegraphics[width=0.45\textwidth]{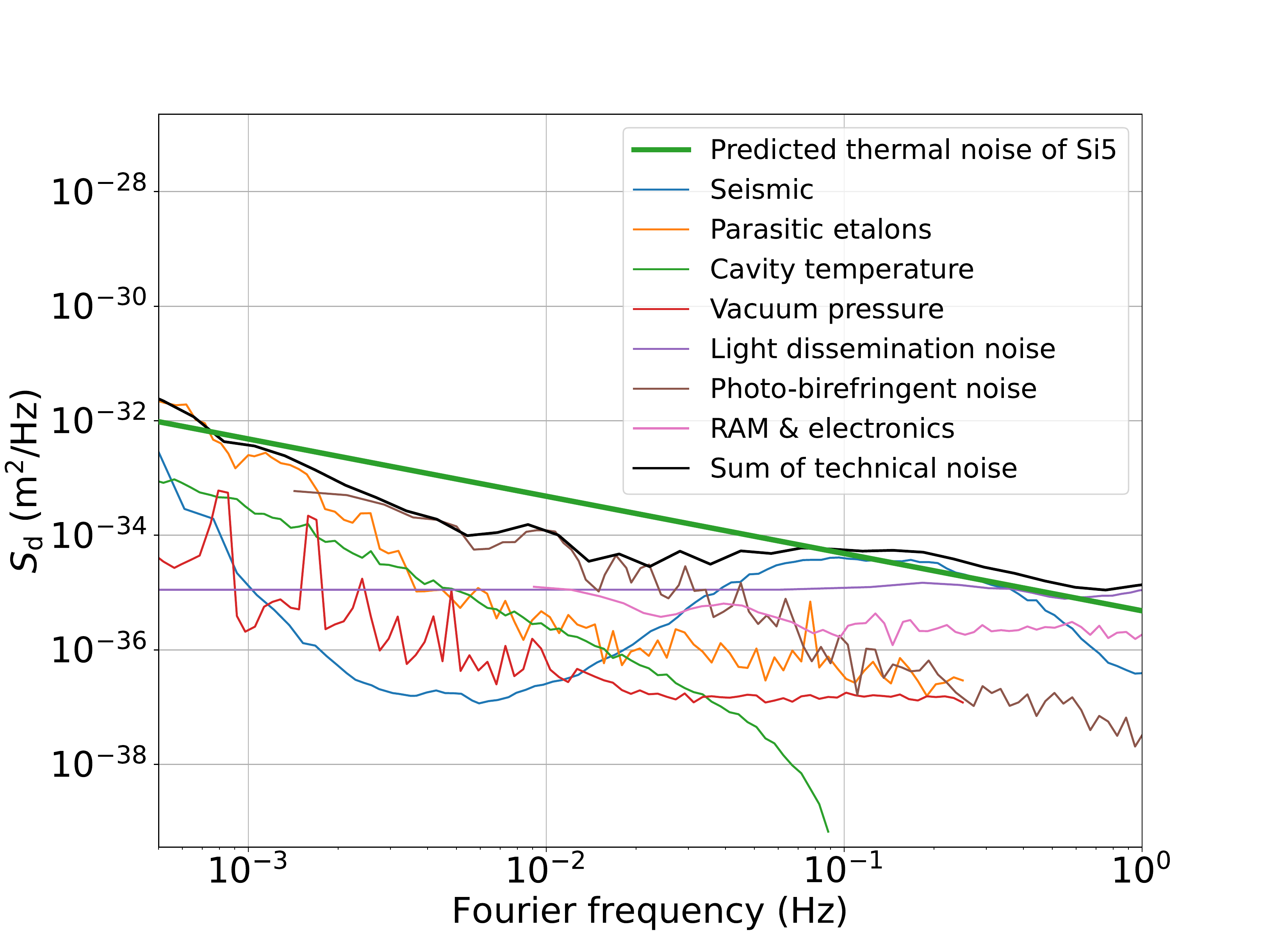}
	\caption{Technical noise contributions for Si5 in comparison to the Brownian noise of the AlGaAs/GaAs coating with $\phi_\mathrm{124K}=2.5\times10^{-5}$.}
    \label{fig:Si5_tech}
\end{figure}

The main technical noise contributions were carefully characterized and minimized as described in Appendix \ref{sec:resonators}. 
The result of the 21~cm silicon resonator is shown in Fig. \ref{fig:Si5_tech}. 
It includes contribution from vibrations calculated from the measured sensitivities and the vibrations at the cavity,
parasitic etalons that were investigated from the observed frequency shifts induced by ambient pressure.
Influence of temperature is estimated from observed temperature fluctuations, including the thermal model and the estimated thermal expansion coefficient of the cavity.
The pressure influence is based on the measured pressure fluctuations and the refractivity of air \cite{bir94}.
Light dissemination noise results from optical path length fluctuations in unstabilized free space and short fiber sections.
The contribution from photo-birefringence is based on the measured sensitivities (see main text) and measured intracavity power fluctuations.
Influence of residual amplitude modulation (RAM) and electronics is based on independently measured error signals.

In the range between 0.75 and 100 mHz, technical noise contributions are mostly below the predicted Brownian thermal noise. 

Thermal noise from the bonding of the crystalline coatings to the silicon mirror substrate should have similar spatial property (local) as the optical coating noise, as this can be approximately treated as an additional layer on the coating with different mechanical loss. 
Therefore, the long spatial coherence length of the excess noise can not be explained by this source.

Brownian noise from the silicon spacer or the mounting would appear as global noise.
This would require an increase of the loss coefficient of silicon by more than an order of magnitude compared to well established values. In addition, there is also no visible difference of the excess noise level between 4 K and 16 K which would appear in this case, as the loss coefficient is constant in this range \cite{naw08}.

\subsection{Instability of the 124 K setup from three-cornered-hat analysis}

The performance of the 21 cm resonator is determined by analyzing the beat signals with two additional oscillators as explained in Appendix \ref{sec:resonators}.
\begin{figure}[!ht]
	\centering\includegraphics[width=0.45\textwidth]{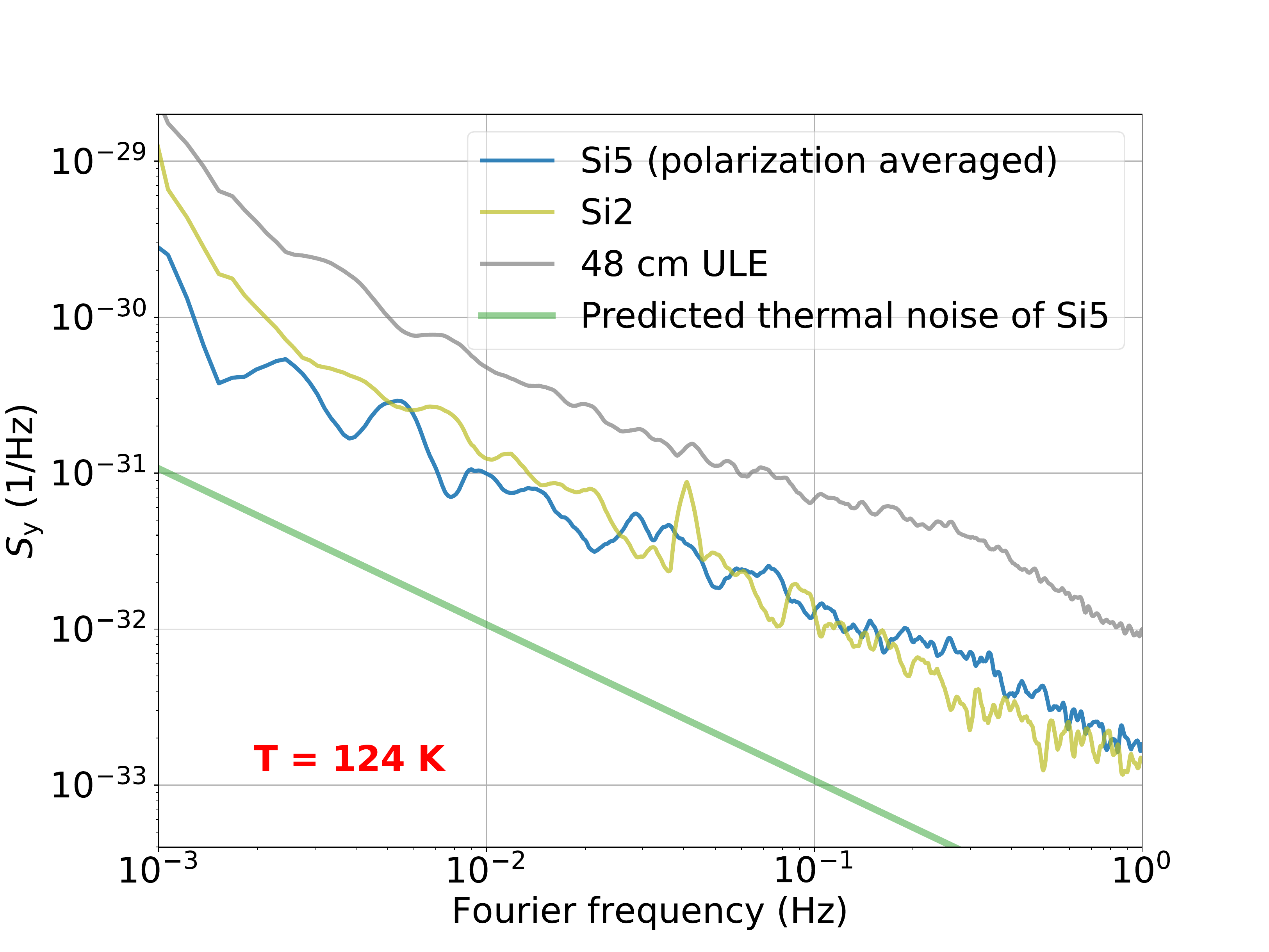}
	\caption{Obtaining by the three-cornered-hat analysis: PSD of fractional frequency noise of Si5 (blue), Si2 (yellow) and the 48 cm ULE cavity (grey). The expected thermal noise of Si5 (green) is included for reference.}
	\label{fig:Si_TCH}	
\end{figure}
The result of this analysis is the power spectral density of fractional frequency fluctuations $S_y(f)$ , which is displayed in Fig.~\ref{fig:Si_TCH}.
The corresponding spectrum of fluctuations $d$ of the total optical length $L_\mathrm{cav}$ between the mirrors is calculated as $S_d = L_\mathrm{cav}^2  S_y$.

The stability of Si5 outperforms Si2 at lower Fourier frequencies, because Si2 suffers from strong parasitic etalons.
Owing to better suppression of parasitic etalons, Si5 has a much better long-term stability than the other two systems.
The performance of the reference system for characterization of Si6 is described in ref.\cite{oel19}.

\section{Averaging birefringent noise - dual frequency lock}\label{sec:2FreqLock}

\begin{figure}[!ht]
	\centering\includegraphics[width=0.45\textwidth]{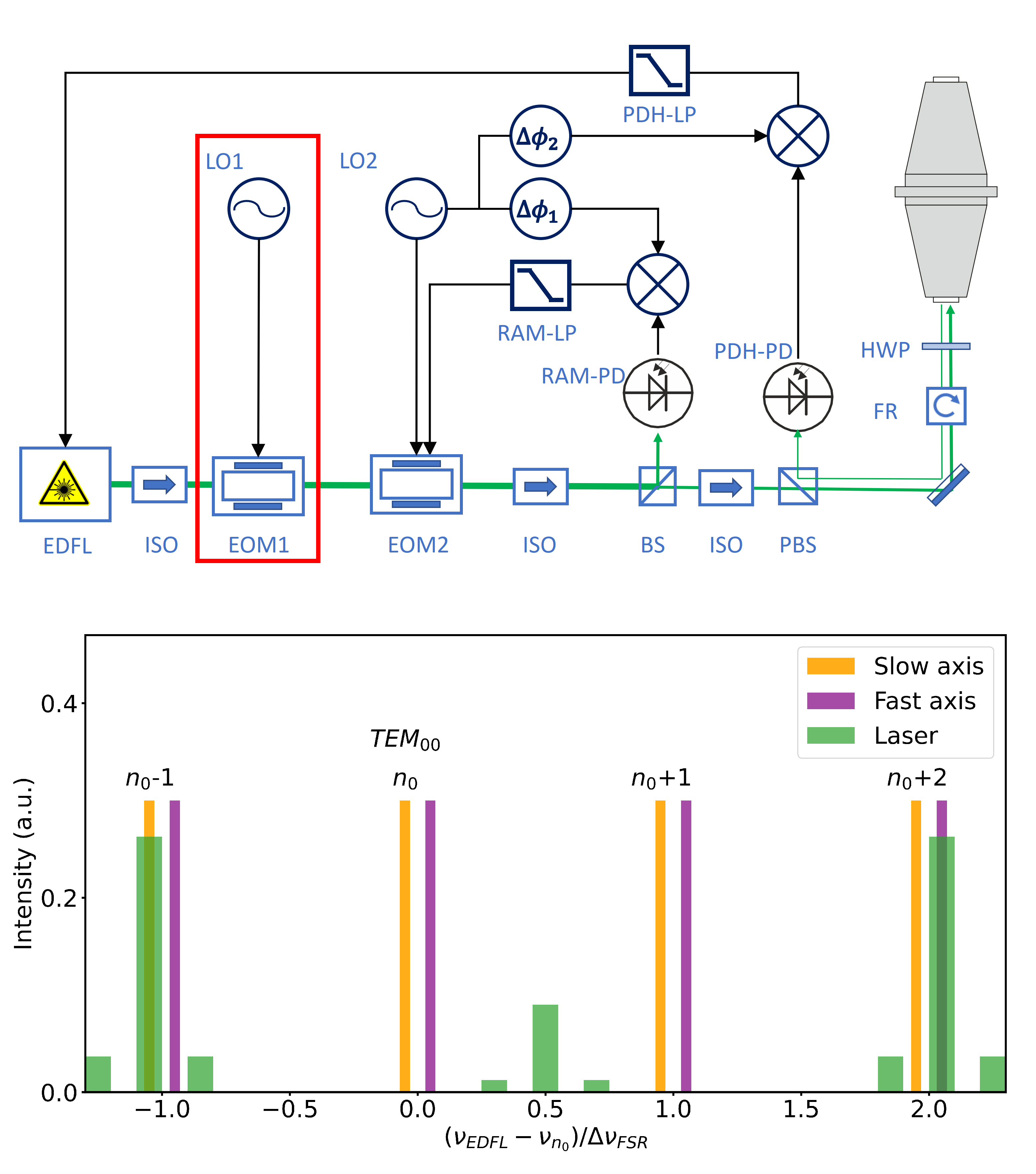}
	\caption{
		(top) 
		Experimental setup of the locking scheme. 
		The additional EOM1 and its driver LO1 required for dual frequency locking are shown in the red box.
		EDFL: erbium doped fiber laser, ISO: optical isolator, EOM: electro-optic modulator, (P)BS: (polarization) beam splitter, 
		FR: Faraday rotator, HWP: half wave plate, PD: photodetector, LP: loop filter, LO: local oscillator.
		(bottom) 
		Frequency components of the light (green), and the cavity resonances of fast (purple) and slow (red) axes in units of the free spectral range $\Delta\nu_\mathrm{FSR}$.
	}
	\label{fig:2FreqLock}
\end{figure}

In this section, we describe the simple dual-frequency locking scheme that was used in our experiment to cancel the anti-correlated birefringent noise.
Using a single laser beam from one side of the cavity, this method generates an error signal with equal contributions from both polarization eigenmodes. 
This technique is much simpler than using two separate lasers from both ends, which requires two independent locking setups including separate RAM control, laser power stabilization, and fiber noise cancellation.
The main building blocks of this locking scheme are shown in Fig. \ref{fig:2FreqLock}. 
The scheme is largely based on the usual PDH setup \cite{dre83} with RAM compensation \cite{won85} and only an additional EOM to generate sidebands (red box in Fig. \ref{fig:2FreqLock}) is required.

In dual frequency locking, the incoupling light at the front mirror is linearly polarized at 45$^\circ$ relative to both axes (fast and slow). 
The first order sidebands of an electro-optic modulator (EOM1 in Fig. \ref{fig:2FreqLock}) are used to excite the two polarization eigenmodes and generate an PDH error signal given by the sum of both modes. 
The modulation index of EOM1 is set to $M = 1.8~\mathrm{rad}$ to maximize the optical power in the two sidebands.
Scanning the laser frequency over the cavity resonance, the PDH-signal shows three components:  when the upper sideband is resonant with the slow axis ($\nu_\mathrm{slow} < \nu_\mathrm{fast}$), when both sidebands are in resonance with their corresponding polarization eigenmodes, and when the lower sideband is resonant with the fast axis. 
By locking to the central error signal, the laser is stabilized to the average of two polarization eigenmodes if the error signals from both modes are equally weighted. 
The RAM control loop stabilizes in this case the RAM of sum of all the spectral lines generated by EOM1, which is dominated by the 1st order sidebands used for dual frequency locking.
Stabilizing the laser using first order sidebands on the two polarization eigenmodes of the same longitudinal mode ($f_\mathrm{mod1} = 0.5 \Delta \nu_\mathrm{birefr}$) is not advisable.
The small separation between the spectral lines would lead to significant coupling of noise in the wings of one sideband to the adjacent polarization eigenmode
and the undesired interference with the other sideband would degrade the frequency stability. 
Thus the modulation frequency $f_\mathrm{mod1}$ of EOM1 is set to address polarization modes separated by at least one free spectral range $\Delta \nu_\mathrm{FSR}$:
\begin{equation}
	f_\mathrm{mod1} = (n + 0.5) \Delta \nu_\mathrm{FSR} \pm 0.5\,\Delta \nu_\mathrm{birefr}, n = 0, 1, 2 ...
\end{equation}

If only one laser is stabilized to the resonator via dual frequency locking, $n$ can be set to 0 ($f_\mathrm{mod1} \approx 0.5~\Delta \nu_\mathrm{FSR}$). 
\begin{figure}[!ht]
	\centering\includegraphics[width=0.45\textwidth]{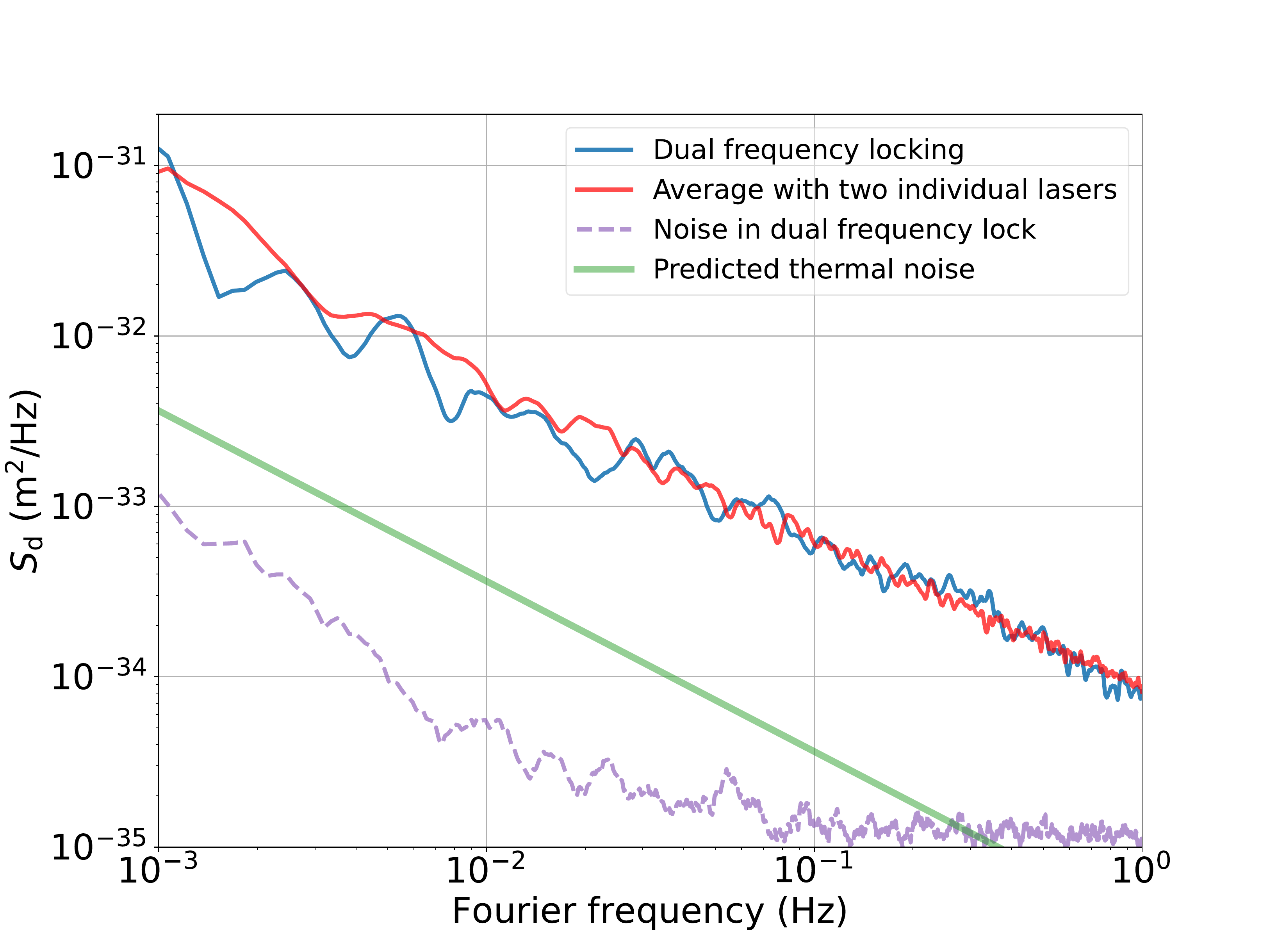}
	\caption{PSD of dual-frequency locking (blue) compared to that achieved by averaging with two independent laser setups. The PSD of the difference between two lasers stabilized to adjacent HG$_{00}$ modes with dual frequency locking indicates the quality of this technique. Both methods suppress the birefringent noise by at least a factor of 10. }
	\label{fig:2FreqLock2}
\end{figure}
To lock two lasers simultaneously with dual frequency locking from opposite sides of the resonator, we choose
$f_\mathrm{mod1} = 1.5\,\Delta \nu_\mathrm{FSR} + 0.5\,\Delta \nu_\mathrm{birefr}$ 
to avoid crosstalk between the two lasers and to simplify the beat detection for frequency counting: 
the smallest beat frequency between the top and bottom lasers is obtained by locking the two lasers to adjacent cavity modes, and by generating a beat between one of the first order sidebands of the bottom laser (transmission) and the off-resonance carrier of the top laser (reflection). 

The performance of the dual frequency lock is reduced by the smaller error signal in comparison to normal PDH-locking, 
and the imperfect weighting of the error signals of the two polarizations. 
The smaller error signal arises from the polarization mismatch between incident light field and the polarization axes of the resonator, as only half of the optical power of the corresponding spectral lines can be coupled into the cavity and contributes to the error signal. 
To balance the error signals from the two polarization eigenmodes, we optimize the settings of the half wave plate in front of the resonator. 
From the comparison with locking two independent lasers and thus perfect averaging (Fig. \ref{fig:2FreqLock2}), a similar suppression of birefringent noise has been achieved which corresponds to a tenfold reduction of birefringent noise in PSD (Fig. \ref{fig:birefr}b).

To evaluate the ultimate noise of this method, we stabilize both lasers via dual frequency locking to adjacent polarization-averaged TEM$_{00}$-modes. The noise between the two lasers  shows a total noise level well below the global excess noise and even below the predicted coating Brownian thermal noise (see Fig. \ref{fig:2FreqLock2}).

\section{Experimental methods and further results}
\subsection{Spatial correlation of Brownian thermal noise}\label{sec:BTN_spatial}
According to the fluctuation dissipation theorem, the PSD of Brownian thermal noise is proportional to the average dissipated power $W_\mathrm{diss}$ when a pressure with shape of beam intensity profile generated by a force $F_0$ oscillating at the corresponding frequency $f$ is applied \cite{lev98,kes12}.
Provided the mode diameter is much larger than the coating thickness, the single-sided PSD of the coating Brownian thermal fluctuations probed by a HG\textsubscript{mn}-mode is:\\
\begin{widetext}
	\begin{equation}
		S_{\mathrm{Brown}}^{(mn)}=\frac{2k_{B}T}{\pi^2f^2}\cdot \frac{W_\mathrm{diss}^{(mn)}}{F_0^2}=g_{(mn)} \cdot \frac{4k_{B}T(1+\sigma_\mathrm{sub})(1-2\sigma_\mathrm{sub})d_\mathrm{ct} }{\pi^{2}fw^{2}E}\phi_{\mathrm{ct}}
	\end{equation}
\end{widetext}
where $k_B$ is the Boltzmann constant,
$T$ is resonator temperature,
$\sigma_{sub}$ and $E$ are the Poisson ratio and Young's modulus of mirror substrate,
$f$ is the Fourier frequency,
$w$ is the $1/e^2$ beam radius on the mirror,
$d_\mathrm{ct}$ and $\phi_{\mathrm{ct}}$ are the thickness and mechanical loss of the coating.
The merit factor $g_{(mn)}=S_{\mathrm{Brown}}^{(mn)}/S_{\mathrm{Brown}}^{(00)}$ was introduced by Vinet et. al. \cite{vin10} to describe the scaling of Brownian thermal noise between HG\textsubscript{00} and HG\textsubscript{mn}-mode:
\begin{widetext}
	\begin{equation}
		g_{(mn)}=\frac{4}{\pi} \int_{0}^{\infty} dp \int_{0}^{\infty} dq \;e^{-(p^2+q^2)} \left(L_{m}(p^2) \times L_{n}(q^2)\right)^2
	\end{equation}
\end{widetext}
where $L_{m}(x)$ is the m-th ordinary Laguerre polynomial. Table \ref{table:g_mn} gives the first $g_{(mn)}$-factors, and the factor relevant for this work is $g_{(00)}/g_{(01)}=1.33$.

The predicted Brownian thermal noise for 21 cm and 6 cm silicon resonators at different operating temperatures are shown in the green curves in Fig. \ref{fig:birefr}b-d.\\
Similarly, the fluctuations of the frequency difference between Hermite-Gaussian modes HG$_{mn}$ and HG$_{00}$ induced by coating Brownian thermal noise $S_{\mathrm{Brown}}^{\Delta(mn)}$ can be calculated by applying a pressure, which has a shape of the intensity profile difference between the two cavity modes. Following the formalismus from Vinet et. al.\cite{vin10}, the scaling factor is:
\begin{widetext}
	\begin{eqnarray}
		&g_{\Delta(mn)}&=S_{\mathrm{Brown}}^{(mn)}/S_{\mathrm{Brown}}^{(00)} \nonumber \\
		&&=\frac{4}{\pi} \int_{0}^{\infty} dp \int_{0}^{\infty} dq \;e^{-(p^2+q^2)} \left(L_{m}(p^2) \times L_{n}(q^2)-L_{0}(p^2) \times L_{0}(q^2)\right)^2
	\end{eqnarray}
\end{widetext}
The numerical values for the first $g_{\Delta(mn)}$-factors  can be found in Table \ref{table:g_delta}, and the factor relevant for this work is $g_{(00)}/g_{\Delta(01)}=1.33$. 
Therefore, from the measured noise in the difference between HG\textsubscript{00} and HG\textsubscript{01}-mode, the local (Brownian) noise for any HG\textsubscript{mn}-mode can be calculated.

The correlation coefficient between the coating Brownian thermal noise of an HG$_{mn}$ mode and the HG$_{00}$ mode $\mathrm{corr}_{(mn)}$ can be calculated as:
\begin{eqnarray}
	\mathrm{Corr}_{(mn)}&=&\frac{S_\mathrm{Brown}^{(00)}+S_\mathrm{Brown}^{(01)}-S_\mathrm{Brown}^{\Delta(mn)}}{2\sqrt{S_\mathrm{Brown}^{(00)}S_\mathrm{Brown}^{(mn)}}}\\
	&=&\frac{1+g_{(mn)}-g_{\Delta(mn)}}{2\sqrt{g_{(mn)}}}
\end{eqnarray}
The numerical values of the first correlation coefficients $\mathrm{Corr}_{(mn)}$ can be found in table \ref{table:corr-ct}.
\begin{table}[ht!]
	\caption{Numerical values for $g_{(mn)}$, the relative PSD of HG$_{mn}$ to HG$_{00}$.}
	\label{table:g_mn}
	\begin{ruledtabular}
		\begin{tabular}{cc|cccc}
			&m	&0	&1	&2	&3	\\
			n&	&	&	&	&	\\
			\hline
			0&	&1.000&0.750&0.641&0.574\\
			1&	&0.750&0.563&0.480&0.431\\
			2&	&0.641&0.480&0.410&0.368\\
			3&	&0.574&0.431&0.368&0.330\\
		\end{tabular}
	\end{ruledtabular}
\end{table}

\begin{table}[ht!]
\caption{Numerical values for $g_{\Delta\left(mn\right)}$, the relative PSD of the difference between HG$_{mn}$ and HG$_{00}$ to HG$_{00}$.}
\label{table:g_delta}
	\begin{ruledtabular}
		\begin{tabular}{cc|cccc}
			&m	&0	&1	&2	&3	\\
			n&	&	&	&	&	\\
			\hline
			0&	&0.000&0.750&0.890&0.949\\
			1&	&0.750&1.063&1.105&1.118\\
			2&	&0.890&1.105&1.129&1.133\\
			3&	&0.949&1.118&1.133&1.134\\
		\end{tabular}
	\end{ruledtabular}
\end{table}

\begin{table}[ht!]
	\caption{Numerical values for the correlation coefficients between Brownian noise of HG$_{mn}$ and HG$_{00}$ mode $\mathrm{Corr}_{(mn)}$ .}
	\label{table:corr-ct}
	\begin{ruledtabular}
		\begin{tabular}{cc|cccc}
			&m	&0	&1	&2	&3	\\
			n&	&	&	&	&	\\
			\hline
			0&	&1.000&0.577&0.469&0.412\\
			1&	&0.577&0.333&0.271&0.238\\
			2&	&0.469&0.271&0.219&0.194\\
			3&	&0.412&0.238&0.194&0.171\\
		\end{tabular}
	\end{ruledtabular}
\end{table}

%
%

\subsection{Spatial property of birefringent noise}\label{sec:BR_spatial}

To investigate the spatial property of the birefringent noise, one needs to compare the noise between fast and slow mode of two different HG-modes, which would require in total four lasers.
To avoid the associated complexity, we analyzed the birefringent noise with a different approach using only two lasers, where we assume that noise level of the polarization independent noise remains constant between different measurements.
For these investigations we stabilize one laser to the fast HG$_{00}$-mode, and the other to the slow HG$_{01}$-mode and record their frequency difference and also the frequency difference of these lasers to two reference lasers (Si2 and the 48 cm ULE cavity).
This allows to determine the noise between the two modes and also the noise of each individual laser from a three cornered hat analysis using the two reference lasers.
The displacement fluctuations of the two HG modes and their difference can be expressed as:
\begin{equation}
	d^{\mathrm{(00)fast}}(t)=d_{\mathrm{Brown}}^{\mathrm{(00)}}(t)+d_{\mathrm{global}}(t) + d_{\mathrm{birefr}}^\mathrm{(00)}(t)
\end{equation}
\begin{equation}
	d^{\mathrm{(01)slow}}(t)=d_{\mathrm{Brown}}^{\mathrm{(01)}}(t)+d_{\mathrm{global}}(t) -  d_{\mathrm{birefr}}^\mathrm{(01)}(t)
\end{equation}
\begin{equation}
	d^{\mathrm{(\Delta)}}(t)=d_{\mathrm{Brown}}^{\mathrm{(00)}}(t)-d_{\mathrm{Brown}}^{\mathrm{(01)}}(t) + d_{\mathrm{birefr}}^{\mathrm{(00)}}(t)-d_{\mathrm{birefr}}^{\mathrm{(01)}}(t)
\end{equation}
The corresponding PSDs, which can be determined experimentally via TCH-analysis, are:
\begin{equation}
	S_d^{\mathrm{(00)fast}}=S_{\mathrm{Brown}}^{\mathrm{(00)}}+S_{\mathrm{global}} + S_{\mathrm{birefr}}^\mathrm{(00)}
	\label{eq:Sd_00}
\end{equation}
\begin{equation}
	S_d^{\mathrm{(01)slow}}=S_{\mathrm{Brown}}^{\mathrm{(01)}}+S_{\mathrm{global}}+  S_{\mathrm{birefr}}^\mathrm{(01)}
	\label{eq:Sd_01}
\end{equation}
\begin{equation}
	S_d^{\mathrm{(\Delta)}}=S_{\mathrm{Brown}}^{\mathrm{(\Delta)}}+S_{\mathrm{birefr}}^{\mathrm{(\Delta)}}.
	\label{eq:Sd_delta}
\end{equation}
The PSD $S_{\mathrm{Brown}}^{\mathrm{(\Delta)}}$ for the polarization averaged difference between the HG modes was separately measured using dual frequency locking to these modes, as well as the  PSDs for the polarization averaged fluctuations 
\begin{equation}
	S_d^\mathrm{(00)avg} = S_{\mathrm{Brown}}^{\mathrm{(00)}}+S_{\mathrm{global}}
\end{equation} 
\begin{equation}
	S_d^\mathrm{(01)avg} = S_{\mathrm{Brown}}^{\mathrm{(01)}}+S_{\mathrm{global}}
\end{equation} 
that were obtained from a TCH-analysis.

\begin{figure}[!ht]
	\centering\includegraphics[width=0.45\textwidth]{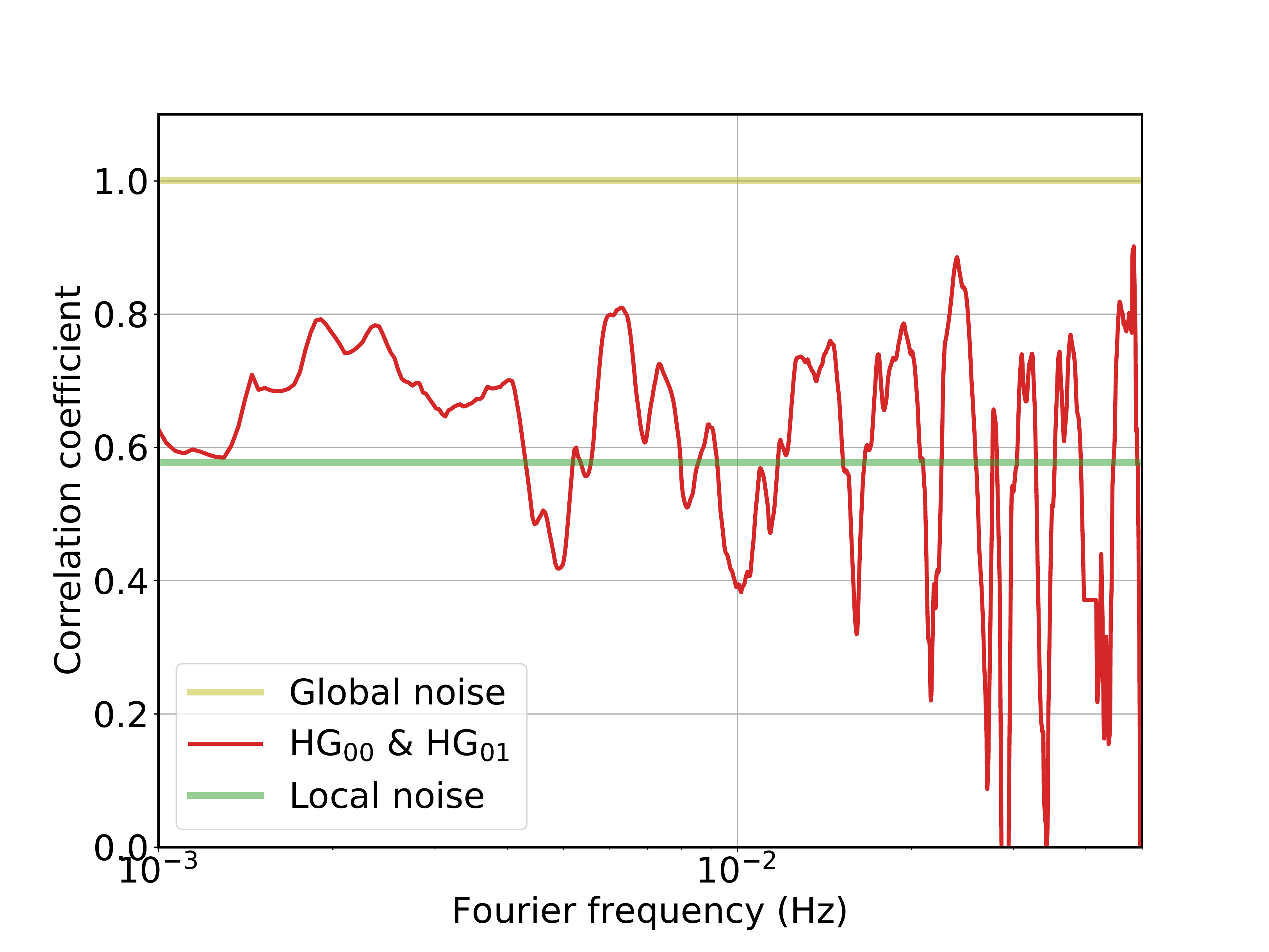}
	\caption{
		Correlation coefficient of birefringent noise between HG$_{00}$ and HG$_{01}$ modes (red). It agrees well with the expectation value for local noise (green), and is significantly different from the expectation value for global noise (yellow). }
	\label{fig:birefr_corr}
\end{figure}

Subtracting these polarization independent PSDs from Eqs. \ref{eq:Sd_01} - \ref{eq:Sd_delta} yields the birefringent noise of the two individual HG-modes and their spatially uncorrelated contribution.
Based on these values, we calculate the correlation coefficient of the birefringent noise:
\begin{equation}
	\mathrm{Corr}=\frac{S_{\mathrm{birefr}}^\mathrm{(00)}+S_{\mathrm{birefr}}^\mathrm{(01)}-S_{\mathrm{birefr}}^{\mathrm{(\Delta)}}}{2\cdot \sqrt{S_{\mathrm{birefr}}^\mathrm{(00)}\cdot S_{\mathrm{birefr}}^\mathrm{(01)}}}
\end{equation}\\
The result $0.59\pm0.18$ agrees with the expected of 0.577 for pure local noise (table \ref{table:corr-ct}) and is well below the global noise (1.00) as shown in Fig. \ref{fig:birefr_corr}, thus indicating the local property of birefringent noise.

\subsection{Power dependence of the photo-birefringent effect}\label{sec:PBR_power}

\begin{figure}[b]
	\centering\includegraphics[width=0.45\textwidth]{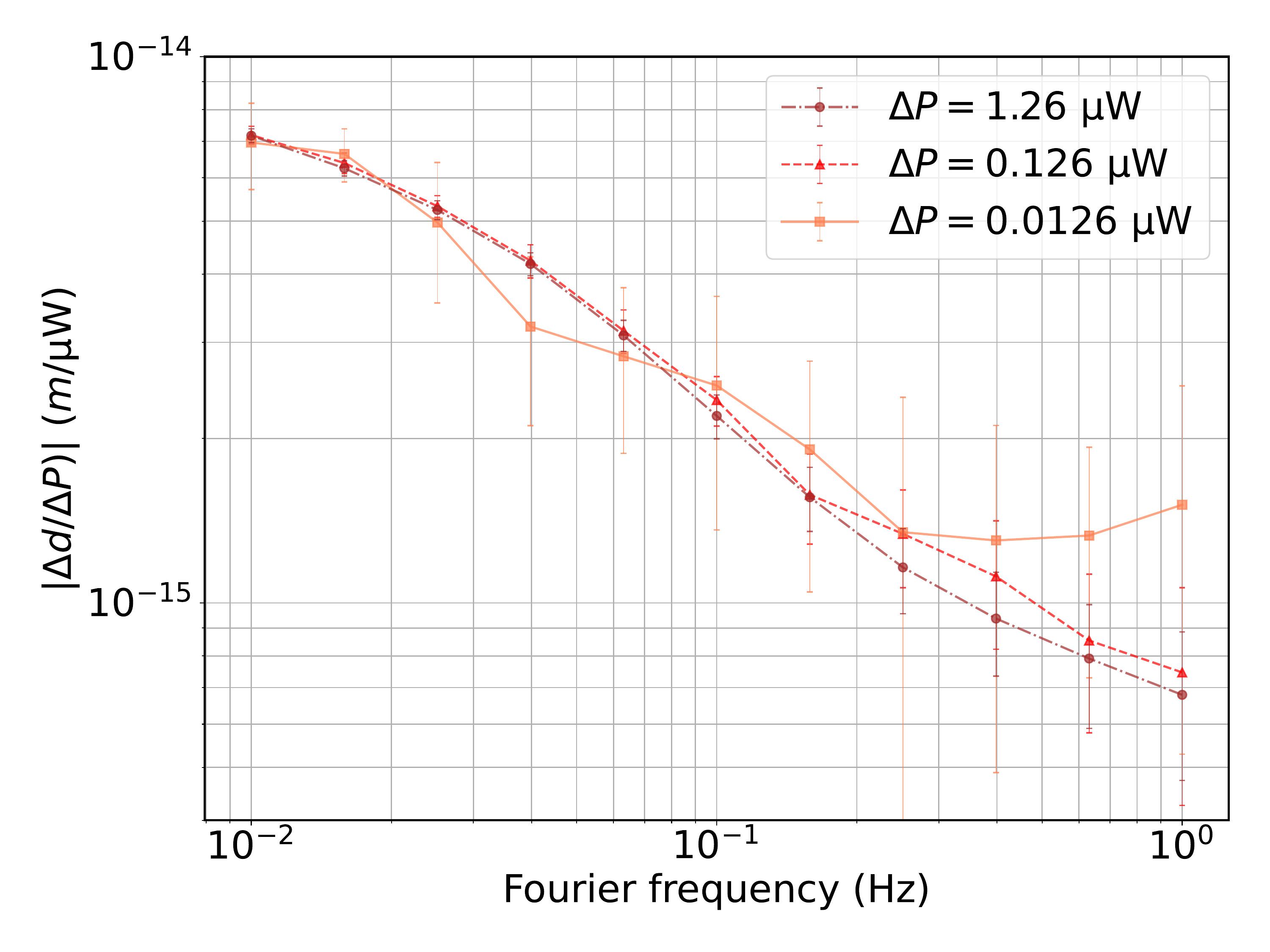}
	\caption{
    Small-signal transfer function measured with different modulation amplitude (slow axis) at 0.62 $\mathrm{\mu W}$/ 1.72 $\mathrm{\mu W}$ mean transmitted power in the fast/slow axis. 
    The error bars indicate the 95\% confidence interval, with contributions from slowly varying electronic offsets, which is estimated by forwarding typical electronic offsets to the slope of PDH error signal. }
	\label{fig:PB_c}
\end{figure}

To make sure that the small-signal transfer function can be applied for our estimation of the photo-birefringent noise due to laser power fluctuations, we determine the transfer function by modulating the optical power coupled to slow axis with different amplitudes $\Delta P$.
The average transmitted optical powers in the fast $P_\mathrm{fast}=0.62~\mathrm{\mu W}$ and in the slow axis $P_\mathrm{slow}=1.72~\mathrm{\mu W}$ are kept constant.
The result (Fig. \ref{fig:PB_c}) shows that a unique transfer function can be used in our noise estimation.

\begin{figure}[!ht]
	\centering\includegraphics[width=0.45\textwidth]{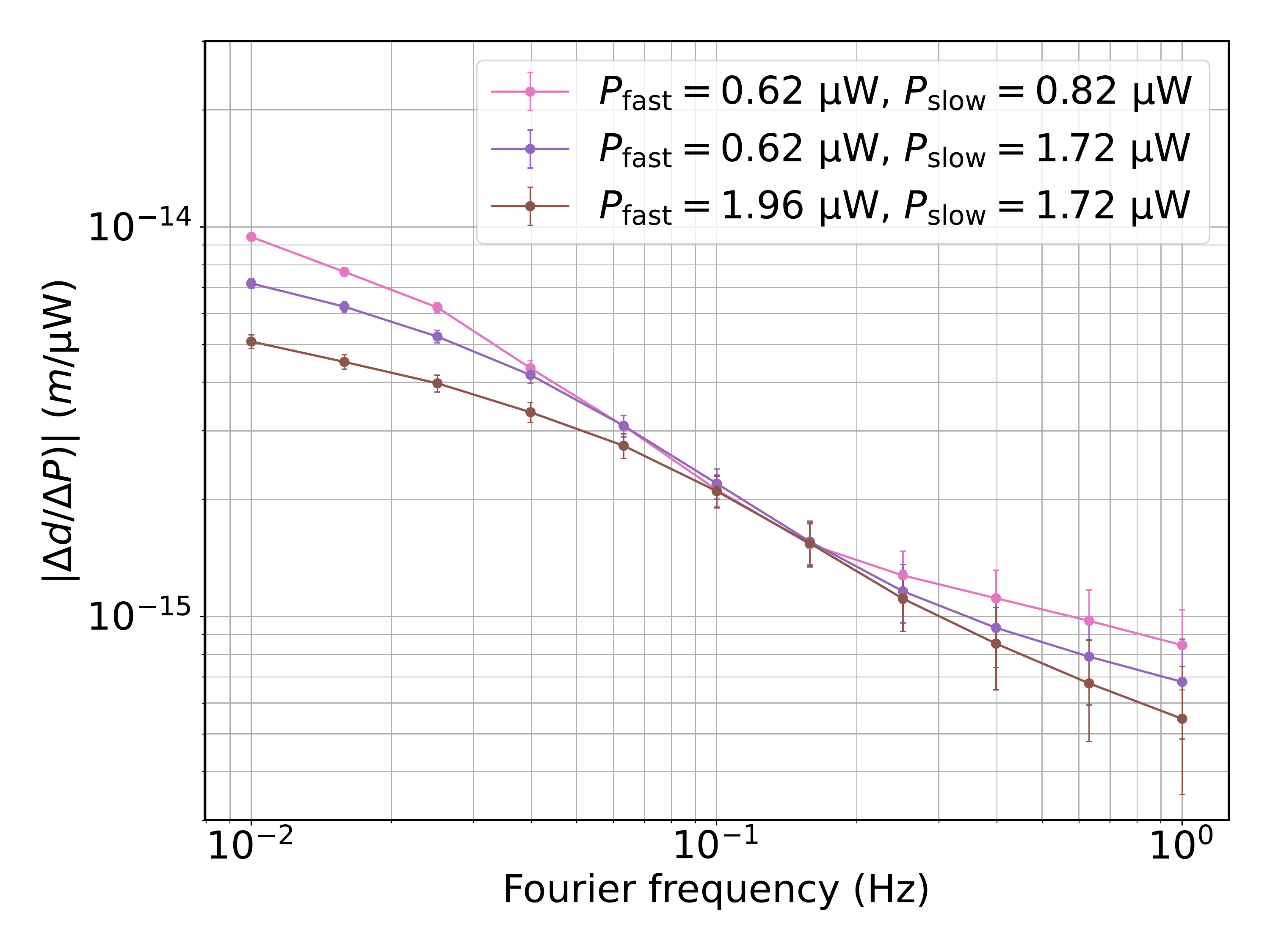}
	\caption{Small-signal transfer function from optical power modulation in transmission $\Delta P$ to optical path length changes $\Delta d$ for different averaged transmitted power $P_\mathrm{trans}$. The error bars have the same meaning as those described in Fig. \ref{fig:PB_c}.}
	\label{fig:PB_b}
\end{figure}

With constant modulation amplitude of $\Delta P~=~0.126~ \mathrm{\mu W}$, the transfer function is measured at three different intracavity power levels (Fig. \ref{fig:PB_b}).
At low Fourier frequencies, the transfer functions depend on the mean transmitted power $P_\mathrm{trans}$, which is proportional to the intracavity power : $P_\mathrm{intra}\approx 2\mathcal{F}/\pi \cdot P_\mathrm{trans}$, where $\mathcal{F}$ is the finesse and assuming the transmission and loss of the coating are equal. 
With the Finesse of $360\,000$ we estimate a factor 
$P_\mathrm{intra}/P_\mathrm{trans} = 0.229~\mathrm{W/\mu W}$.
At higher frequencies the uncertainties are larger and the differences are not significant.

The uncertainty of the transfer function includes statistical uncertainties and contributions from electronic offsets in our measurement.
Electronic offsets $\Delta V$ arising e.g. from the RF-mixer and RF pickup in the Pound-Drever-Hall (PDH) error signal shift the locking frequency away from the cavity resonance by $\Delta\nu = \Delta V/D$, where $D$ is the slope of PDH error signal that is proportional to the optical power.
Hence, modulating the optical power changes the frequency offset $\Delta \nu$ in the PDH lock and adds uncertainty to the measurement.
In addition, electronic offset in the control loop for compensation of residual amplitude modulation (RAM) in combination with the low corner frequency (0.03 Hz) of the RAM loop filter further contributes to the total uncertainty.
Using typical values for offsets in these loops we arrive at the uncertainties shown in Fig. \ref{fig:PB_b} and in Fig. \ref{fig:transient_b}.

\section{Estimated influence on the Einstein-Telescope}\label{sec:ET}

The low frequency Einstein-Telescope (ET-LF) is currently designed to be operated at 10 - 20 K and 1.5 $\mathrm{\mu m}$ wavelength.
One of the candidate test masses is single crystal silicon mirror substrate with crystalline mirror coatings \cite{ET20}.
The detector has a design beam radius of $w_\mathrm{ET}=9$ cm to reduce the coating Brownian noise well below the quantum noise at an intracavity power level of $P_\mathrm{ET}~=~18$ kW, corresponds to a light intensity of $I_\mathrm{ET}\approx71~\mathrm{W/cm^2}$.
The $L_\mathrm{arm}=10$ km arm length of the interferometers further lowers the influence of known coating noise and quantum noise on strain sensitivity to achieve the design value of $h\approx 1 \times 10^{-24}\; \mathrm{/\sqrt{Hz}}$ at $f~=~$10 Hz Fourier frequency.
This corresponds to the temperature and light intensity that are quite similar to the conditions in our Si6 system.
To meet the scientific goal of ET-LF \cite{ET20}, the total coating noise PSD must lower than a quarter of the Brownian thermal noise of the dielectric coating at the operation temperature 
$S_\mathrm{Brown}= 4.5\times10^{-40}~\mathrm{m^2/Hz} \cdot (f/\mathrm{Hz})^{-1}$, 
which corresponds to a noise level of 
$S_\mathrm{ET}(\mathrm{10~Hz}) = 0.25~S_\mathrm{Brown}(\mathrm{10~Hz}) = 1.2\times 10^{-41}~\mathrm{m^2/Hz}$.

From our measurement at 16 K, we observe a birefringent noise level of 
\begin{equation}
 S_\mathrm{birefr}(f) = 2\times 10^{-34}~\mathrm{m^2/Hz}\cdot (f/\mathrm{Hz})^{-1.5}
\end{equation}
with a mode radius of $w_\mathrm{Si6}~=~290~\mathrm{\mu m}$ at $P_\mathrm{Si6}$= 0.15~W intracavity power, which corresponds to a light intensity of $I_\mathrm{Si6}=57~\mathrm{W/cm^2}$.
We verified the local property of the birefringent noise which indicates a $1/w^2$ scaling of the noise PSD.
Furthermore, we observed an empirical scaling of the birefringent noise by $\sqrt{I}$ \cite{ked23}.
Therefore, we estimate that the birefringent noise for ET-LF as:
\begin{eqnarray}
	&S'_\mathrm{birefr} (10~\mathrm{Hz})
 &= S_\mathrm{birefr}(\mathrm{10~Hz})\cdot\sqrt{\frac{I_\mathrm{ET}}{I_\mathrm{Si6}}}\cdot\left(\frac{w_\mathrm{Si6}}{w_\mathrm{ET}}\right)^2  \nonumber\\
 &&=7.3\times 10^{-41}~\mathrm{\frac{m^2}{Hz}},
\end{eqnarray}
which is 6 times higher than $S_\mathrm{ET}$ and 1.6 times that of using dielectric coatings.
Therefore, if the cancellation of birefringent noise is impossible in ET-LF, conventional dielectric coatings will outperform the AlGaAs crystalline coating.

When the birefringent noise is canceled, the global excess noise with large correlation length is the dominating noise contribution with a PSD of 
\begin{equation}
S_\mathrm{global}=3\times10^{-35} \mathrm{m^2/Hz} \cdot (f/\mathrm{Hz})^{-1} 
\end{equation}
in Si6 at 16 K.
If the coherence length $l_\mathrm{corr}$ of $S_\mathrm{global}$ is larger than the beam diameter $2w_\mathrm{ET}=18$~cm, the global excess noise level at 10 Hz would be
\begin{equation}
S'_\mathrm{global}(\mathrm{10~Hz})=3\times10^{-36}~\mathrm{\frac{m^2}{Hz}}.
\end{equation}
If the coherence length is on the order of $l_\mathrm{corr}~=~$1 mm - the lower limit deduced from our measurement - the global excess noise would be reduced to:
\begin{eqnarray}
	&S'_\mathrm{global}(10~\mathrm{Hz})
 &= S_\mathrm{global}(\mathrm{10~Hz})\cdot\left(\frac{l_\mathrm{corr}}{2w_\mathrm{ET}}\right)^2 \nonumber\\
 &&=9.3 \times 10^{-41}~\mathrm{\frac{m^2}{Hz}},
\end{eqnarray}
which is still 8 times higher than the ET-LF design value $S_\mathrm{ET}$, and 2.1 times that of using dielectric coatings.

In either case, the current crystalline coating does not meet the requirement of ET-LF.

\begin{widetext}

%


\end{widetext}	
\end{document}